\documentclass[twocolumn]{aastex631}
\usepackage[utf8]{inputenc}
\usepackage{graphicx}
\usepackage{amsbsy}
\usepackage{hyperref}
\usepackage{textgreek}
\usepackage[varg]{txfonts}
\usepackage{svg}

\begin{document}

\title{Interstellar Polarization Survey II: General Interstellar Medium}

\author[0000-0003-0400-8846]{M.J.F. Versteeg}
\affiliation{Department of Astrophysics/IMAPP, Radboud University, PO Box 9010,
6500 GL Nijmegen, The Netherlands}

\author[0000-0002-1580-0583]{A.M. Magalh{\~a}es}
\affiliation{Depto. de Astronomia, IAG, Universidade de São Paulo, Brazil}

\author[0000-0002-5288-312X]{M. Haverkorn}
\affiliation{Department of Astrophysics/IMAPP, Radboud University, PO Box 9010,
6500 GL Nijmegen, The Netherlands}

\author[0000-0001-5016-5645]{Y. Angarita}
\affiliation{Department of Astrophysics/IMAPP, Radboud University, PO Box 9010,
6500 GL Nijmegen, The Netherlands}

\author[0000-0002-9459-043X]{C.V. Rodrigues}
\affiliation{Divis\~ao de Astrof\'\i sica,
Instituto Nacional de Pesquisas Espaciais (INPE/MCTI),
Av. dos Astronautas, 1758,
S\~ao Jos\'e dos Campos, SP, Brazil}

\author[0000-0001-6880-4468]{R. Santos-Lima}
\affiliation{Depto. de Astronomia, IAG, Universidade de São Paulo, Brazil}

\author[0000-0001-6099-9539]{Koji S. Kawabata}
\affiliation{Hiroshima Astrophysical Science Center, Hiroshima University,
Kagamiyama, Higashi-Hiroshima, Hiroshima, 739-8526, Japan}

\begin{abstract}
\edit2{Magnetic fields permeate the entire Galaxy and are essential to, for example, the regulation of several stages of the star formation process and cosmic ray transportation. Unraveling its properties, such as intensity and topology, is an observational challenge that requires combining different and complementary techniques. The polarization of starlight due to the absorption by field-aligned non-spherical dust grains provides a unique source of information about the interstellar magnetic field in the optical band. This work introduces a first analysis } of a new catalog of optical observations of linearly polarized starlight in the diffuse interstellar medium (ISM), the Interstellar Polarization Survey, General ISM (IPS-GI).
We used data from the IPS-GI, focusing on 38 fields sampling lines of sight in the diffuse medium. The fields are about 0.3$^{\circ}$ by 0.3$^{\circ}$ in size and each of them contains $\sim1000$ stars on average. The IPS-GI catalog has polarimetric measurements of \edit2{over 40000} stars, over \edit2{18000} of which have ${P}/\sigma_{P} > 5$. We added distances and other parameters from auxiliary catalogs to over \edit2{36000} of these stars. We analyzed parameter distributions and correlations between parameters of a high-quality subsample of \edit2{10516} stars (i.e. $\sim275$ stars per field).
As expected, the degree of polarization tends to increase with the extinction, producing higher values of polarization at greater distances or at lower absolute Galactic latitudes. Furthermore, we \edit3{find evidence for a} large-scale ordered Galactic magnetic field.
\end{abstract}


\section{Introduction}
\label{intro}
It has been recognized that the magnetic fields in galaxies significantly affect the interstellar medium (ISM). Magnetic fields play a fundamental role in how stars form \edit2{ \citep[][]{1978ppim.book.....S, kulsrud2008origin, elmegreen2004InterstellarTurbulence, Santos-Lima2010DiffusionOfMagneticField, federrath2012sfrturbulentmagnetizedclouds}}. The magnetic field controls the origin and the confinement of cosmic rays in galaxies, high energy particles that play important roles themselves in the galactic environment \edit2{ \citep[][]{skilling1971cosmicrays, casse2001transportcosmicrays,schlickeiser2002cosmicrayastrophysics}}. In addition, magnetic fields are an important component of the energy balance of the Galactic ecosystem \edit2{\citep[][]{ferriere2001interstellar}}.
\par
Light emitted by stars can become polarized as it travels through the ISM \citep{hiltner1949presence, hiltner1949polarization, hall1949Sci...109..166H}. This is caused by interstellar anisotropic extinction. Non-spherical dust grains embedded in the ISM align their short axes with the Galactic Magnetic Field (GMF), polarizing optical starlight parallel to the magnetic field, see for example \citet{andersson2015interstellar} for a recent review. By studying the resulting polarized starlight, we can learn about the dust properties and distribution, as well as the magnetic field structure between the star and observer. More specifically, optical starlight polarization traces the line-of-sight average of the plane-of-sky component of the magnetic field, weighted by the local dust density. Although the observed polarization is the integrated polarization along the whole line of sight to the star, by observing many stars and taking the distance to each of the stars into account, the GMF can be studied in 3 dimensions. 
\par
Optical starlight polarization has been used to study magnetic fields of distinct objects and dense regions of the sky, see for example \citet{crutcher2012magnetic, medhi2008optical, pereyra2007polarimetry}. The largest currently available optical polarization catalog, containing 9286 stars, is an agglomeration of many smaller catalogs \citep{heiles20009286}. The \edit1{Interstellar Polarization Survey (IPS)} program\footnote{\edit1{Based on observations made at the Observatório do Pico dos Dias/LNA (Brazil).}} \citep[][Magalh\~aes et al. in prep.]{magalhaes2005southern} is an effort to increase the number and density of these observations.

\edit2{The present work aims to (i) introduce and describe this new catalog, and (ii) present a first general analysis of these data. In this first study, we explore the correlation of the polarization with different parameters, using data from the \edit1{IPS} program \citep{magalhaes2005southern} and focusing strictly on lines-of-sight in the general ISM (IPS-GI). A more detailed analysis of the structure of the ISM magnetic field based on this survey is forthcoming.}

\par
\edit2{This work is organized as follows.} Section \ref{sec:obs} outlines the survey and describes the acquisition and processing of the data. We cross-correlate our data with auxiliary catalogs in Section \ref{sec:cc}. Section \ref{sec:dataver} describes the verification of the data, including a comparison to the \citet{heiles20009286} catalog. Section \ref{sec:statprop} explores various parameter distributions (spatial, photometric and polarimetric) of the stars. Section \ref{sec:corr} discusses correlations between stellar parameters. Finally, we discuss our findings in Section \ref{sec:disc}, which are summarized in Section \ref{sec:conc}.

\section{Observations}
\label{sec:obs}
\subsection{Data acquisition \label{dataacq}}

The optical polarization data were obtained as a part of the Interstellar Polarization Survey (IPS) program \citep[see][Magalh\~aes et al. in prep.]{magalhaes2005southern}, based on observations made at the Observatório do Pico dos Dias/LNA (Brazil). The polarimeter, a modified CCD camera, consists of a Savart prism and a rotating half-wave waveplate (see \citet{magalhaes1996high} for more details). A remarkable feature of the polarimeter is the simultaneous imaging of both the ordinary and the extraordinary beams, which allows for photon-noise limited observations even under nonphotometric conditions, as well as cancellation of any sky polarization. The polarimeter was mounted onto \edit1{the Cassegrain focus of the IAG Boller \& Chivens 61 cm telescope} at Observatorio Pico dos Dias (OPD), operated by Laboratório Nacional de Astrofísica (LNA) in Brazil. IPS covered different regions of the sky aiming at various distinct science goals such as open clusters or dense clouds. 
\par
\edit1{
The Savart plate allows us to simultaneously image two perpendicularly polarized images of each star in the field. As detailed in \citet[]{magalhaes1984photoelectric, magalhaes1996high}, we can write the ratio between the difference and the sum of these images counts for each object in terms of the Stokes parameters Q and U of the object. This difference to sum ratio varies sinusoidally as a function of the waveplate position.}
\edit1{
Typically, a polarization measurement consists of images taken at 8 waveplate positions (0, 22.5, 45, 67.5, 180, 202.5, 225 and 247.5 degrees) separated by 22.5 degrees. Four of these positions measure alternately Q and -Q (0, 45, 180 and 225 deg) and the other four measure U and -U. This averages out any possible effect from inhomogeneities of the waveplate and also removes any contribution of the background to the polarization.
}
\edit1{
The Stokes parameters are then estimated by a least-square solution of the sinusoidal curve, whose amplitude is defined by Q and U, through the diff/sum points as a function of waveplate position. The accuracy of Q and U (and hence P) are then estimated from the squared differences from each diff/sum point to the fitted curve. One can compare the accuracies thus obtained (and which are the ones quoted in our results) with photon noise estimates as discussed in detail in \citet[in prep.]{magalhaes1984photoelectric, magalhaes1996high}. The accuracies are compatible with the expected photon noise values.
}
\par
For this paper, we looked at the 38 ``General ISM" fields of view (IPS-GI) observed in V band, for which the observations were carried out between April 2000 and August 2003. The Galactic coordinates and number of objects (after applying quality filters, see Section \ref{appfilters}) of these fields can be found in Table~\ref{tab:perfield}. The fields are approximately 0.3 x 0.3 degrees in size and mostly located close to the Galactic plane ($|b| < 10 \degr$). \edit2{For a total of 41108 stars, }ach field contains, on average, about 1000 stars. The fields are primarily centered on stars in the \citet{heiles20009286} catalog. Each field was selected carefully to avoid dense structures and clouds, thus focusing on the diffuse ISM\edit1{.} Some fields were observed more than once using different exposure times, typically ranging from 10 to 30 seconds per waveplate position. The different exposure times ensure a good covering of a wide magnitude range, achieving high signal-to-noise ratio measurements of $P/ \sigma_{P} > 5$ for some faint stars of about $V_{mag} = 18$, although due to the different exposure times we cannot claim completeness. 

\subsection{Data processing \label{dataproc}}
The data was reduced using the SOLVEPOL pipeline, developed specifically for imaging polarimetry data of this type \citep{ramirez2017solvepol}. SOLVEPOL uses the astrometry.net software \citep[][]{lang2010astrometry} to calibrate the astrometry and the Guide Star Catalog version 2.3 \citep[][]{lasker2008second} for the magnitude calibration. Images were corrected for bias and flat-field whenever possible and if no bias frames and flat-fields were available, the raw images were used in the processing. This did not lead to significantly lower quality results. SOLVEPOL produces tables with polarimetric (degree of linear polarization P, polarization angle $\theta$) and photometric (V-band magnitudes) measurements and their associated uncertainties for each detected star. SOLVEPOL uses the following equations to calculate the degree of polarization and the polarization angle (see \cite{ramirez2017solvepol} for more details):

\begin{equation} P = \sqrt{Q^{2}+U^{2}} ,\end{equation}

\begin{equation} \theta = \frac{1}{2}\arctan\frac{U}{Q} ,\end{equation}

\noindent in which Q and U are the normalized Stokes' parameters and P refers to the degree of linear polarization as a fraction of the total intensity. We follow the conventions from \cite{ramirez2017solvepol} and refer to that paper for more details. Degrees of polarization will be presented as a percentage throughout \edit1{the present work}.
\par
The output tables were used for further processing. Entries from stars that were observed more than once were merged. The final value is an average of all observations, weighted with the inverse of the squared uncertainties in the Stokes parameters. Observations of standard stars of known polarization were used to determine the correction of the polarization angle to the equatorial coordinate system. \edit1{To find the instrumental polarization, we used observations of known unpolarized standard stars. We calculated weighted averages of the Stokes parameters Q and U per observing run.} The weights were set equal to the inverse of the squares of the uncertainties in Q and U. The Stokes parameters were then used to find the polarization, following the equations above. This led to an average instrumental polarization of 0.07\%. This is smaller than the observational uncertainties, as the average observational error in the degree of polarization across the entire data set is approximately \edit2{0.3\%}, and thus we did not correct for instrumental polarization. 
\par
In this paper, the averages of the degree of linear polarization P and polarization angle $\theta$ are calculated following the conventions from \citet{pereyra2007polarimetry}. The polarimetric averages are weighted by the inverse of the squared uncertainty in the Stokes parameters Q and U.

\subsection{Debiased Degree of Polarization \label{data:debias}}
Measurements of the degree of linear polarization are biased at low ${P}/\sigma_{P}$ \citep{clarke1986statistical}. The polarization values in this paper have not been debiased. We instead filtered the data to only include sources with a ${P}/\sigma_{P} > 5$, thus avoiding the need to debias data. However, for verification, we compared our biased degrees of polarization to values that have been debiased using the estimator from \citet{plaszczynski2014novel}:

\begin{equation}
P_{d} = P - \sigma_{P}^{2} \frac{1-e^\frac{-P^{2}}{\sigma_{P}^{2}}}{2P},
\end{equation}

\noindent
where $P_{d}$ is the debiased polarization. Correcting the polarization degrees using this estimator did not lead to significant differences in the values. Differences between the biased and debiased polarization values are of the order $10^{-3} \%$, well below the typical uncertainties. This is consistent with the findings of \citet{plaszczynski2014novel}, who find that their estimator becomes indistinguishable from other estimators beyond ${P}/\sigma_{P} > 3$. We therefore conclude the ${P}/\sigma_{P}$ filter to be an effective limiting criterion in avoiding statistical difficulty. Applied to the full, unfiltered dataset, the difference between the biased and debiased values is larger (\edit2{of the order $0.1 \%$ on average, the same order of magnitude as the measurement error}). This shows the need for debiasing if the data were used without stringent ${P}/\sigma_{P}$ filters. We note that the applied ${P}/\sigma_{P}$ filter may inadvertently remove lower polarization stars.

\section{Auxiliary catalogs}
\label{sec:cc}
\subsection{Distances \label{cc:dists}}
Our analysis required each star to have an estimate of its distance and interstellar extinction in addition to its interstellar polarization. Therefore, we cross-correlated our initial polarimetric sample with other catalogs as described below. Each observed field was assigned a unique identifier and each star in the catalog was matched to a GAIA \edit2{EDR3 counterpart \citep{gaia2021edr3} }using \textsc{Topcat} \citep{taylor2005topcat}. Matches are based on position (a 3" margin, corresponding to $1.5 \sigma_{position}$ for IPS) and magnitude (2 mag margin). This margin in magnitude is to account for differences in GAIA's G-band magnitude and IPS' V-band magnitude. Visual inspection of the positions of the stars in each catalog showed that there is little chance of mismatches-- most stars are close to only one other potential match. Even if more than one potential match were nearby, inclusion of the magnitude filter \edit2{often} finds the best match. \edit2{However, uncertain matches will be excluded from further analysis, see also Section \ref{appfilters}. } For \edit2{96\%} of the IPS sources, a GAIA counterpart was found within the given margins. In addition, we added data from the \citet{anders2022photo} \textsc{StarHorse}-based catalog, which includes not only distances, but also V-band extinctions and many other stellar parameters. \citet{anders2022photo} parameters were matched to each star using the unique GAIA EDR3 id. 
\par

\edit2{Another auxialiary catalog that provides distances is that of \citet[][]{bailer2021estimating}, which is also based on GAIA EDR3. However, because we will also be needing other stellar parameters such as extinction, considering the close relationship between extinction and distance, we have decided to use the \citet[][]{anders2022photo} catalog. However, for completeness we include \citet[][]{bailer2021estimating} photogeometric distances in the analysis where appropriate to confirm that the differences between the two catalogs have no effect on our conclusions.
}

\subsection{V-band extinction \label{cc:extinction}}
\edit2{
Considering the close relation between polarization and the distribution of dust, we must critically assess the quality of the \textit{V}-band extinction, as $A_V$ is one of the most direct tracers of dust available in the \cite{anders2022photo} catalog (the median value is called $A_{V50}$ therein). There are multiple reasons for using parameters from \cite{anders2022photo}. Firstly, the Bayesian algorithm, \texttt{StarHorse}, uses the photometry of multiple surveys (e.g.\ \textit{Gaia}-EDR3, 2MASS, AllWISE, PanSTARRS1 DR1, and SkyMapper DR2), and precise parallax observations from \textit{Gaia}-EDR3 \citep{gaia2021edr3}, covering the entire sky and increasing the accuracy of its parameters in comparison with past versions of the catalog. Secondly, due to the location of the IPS-GI fields in the Southern sky, some fields-of-view are either covered poorly, or not at all by other available dust maps such as \cite{marshall2006modelling}, \cite{green20193d}, and \cite{Lallement_2019}. This also includes coverage and precision in distance. Finally, the small size of the fields challenges the resolution of dust maps such as \cite{planck2016galdustemission}; a single pixel, to which a single $A_V$ value is assigned, may contain multiple IPS-GI stars. Additionally, \cite{planck2016galdustemission} measures the total extinction in emission integrated along the entire Galaxy pathlength with no regard for variations along the line-of-sight.  }

\edit2{
We quantitatively compared the $A_{V50}$ values of \cite{anders2022photo} to other available extinction measurements from the \cite{marshall2006modelling} and \cite{green20193d} three-dimensional dust maps. To this end, the \cite{marshall2006modelling} \textit{K}-band extinctions were converted to the \textit{V}-band with the relative extinction value $A_K/A_V = 0.078$ (\citealt{wang2019optical}, table 3). \cite{green20193d} extinctions, which are initially in arbitrary units, were converted to the \textit{V}-band using the method suggested by the author for \texttt{StarHorse}-like data (Equations 30 and 31 from \cite{green20193d}). We found that despite the differences between the methods to map the dust extinction and the limitations on distance and sky coverage, the extinction values of the matching sources are in good agreement with each other. The median systematic difference of \cite{green20193d} and \cite{marshall2006modelling} with \cite{anders2022photo} extinctions is consistent within the $68\%$ confidence interval (between the \textit{16th} and \textit{84th} percentiles of the difference) out to at least $A_{V50} \sim 4$ mag. The median systematic difference increases for $A_{V50} > 4$ mag. A comparison with \cite{planck2016galdustemission} measurements in intermediate latitude IPS-GI fields ($b>7\degr$), showed a similar result. We refer the reader to Angarita et al. in prep. for a continuation of the analysis of the extinctions in the IPS-GI fields. }

In addition to the parameters from IPS-GI, \edit2{\textit{Gaia}-EDR3 and \citet{anders2022photo}}, we also used the \textit{V}-band extinction, $A_{\rm{V50}}$, to derive values for the reddening E(B-V) and hydrogen column density N\textsubscript{H} using the following equations:

\begin{equation} E(B-V) = \frac{A_{V50}}{3.1} ,\end{equation}

\begin{equation} N_\mathrm{H} = 5.8\times 10^{21} cm^{-2} E(B-V), \end{equation}

\noindent \edit2{the latter from \citet[][]{bohlin1978survey}. We will refer to the median (50th percentile) A\textsubscript{V50} values from \citet{anders2022photo} as A\textsubscript{V} henceforth. Similarly, we will use \emph{dist} or \emph{d} for the median (50th percentile) distance $dist_{\rm{V50}}$ from \citet{anders2022photo}. }

\section{Data verification}
\label{sec:dataver}
\subsection{Comparison to Heiles' catalog \label{hcomp}}
\begin{figure*}[ht!]
\plottwo{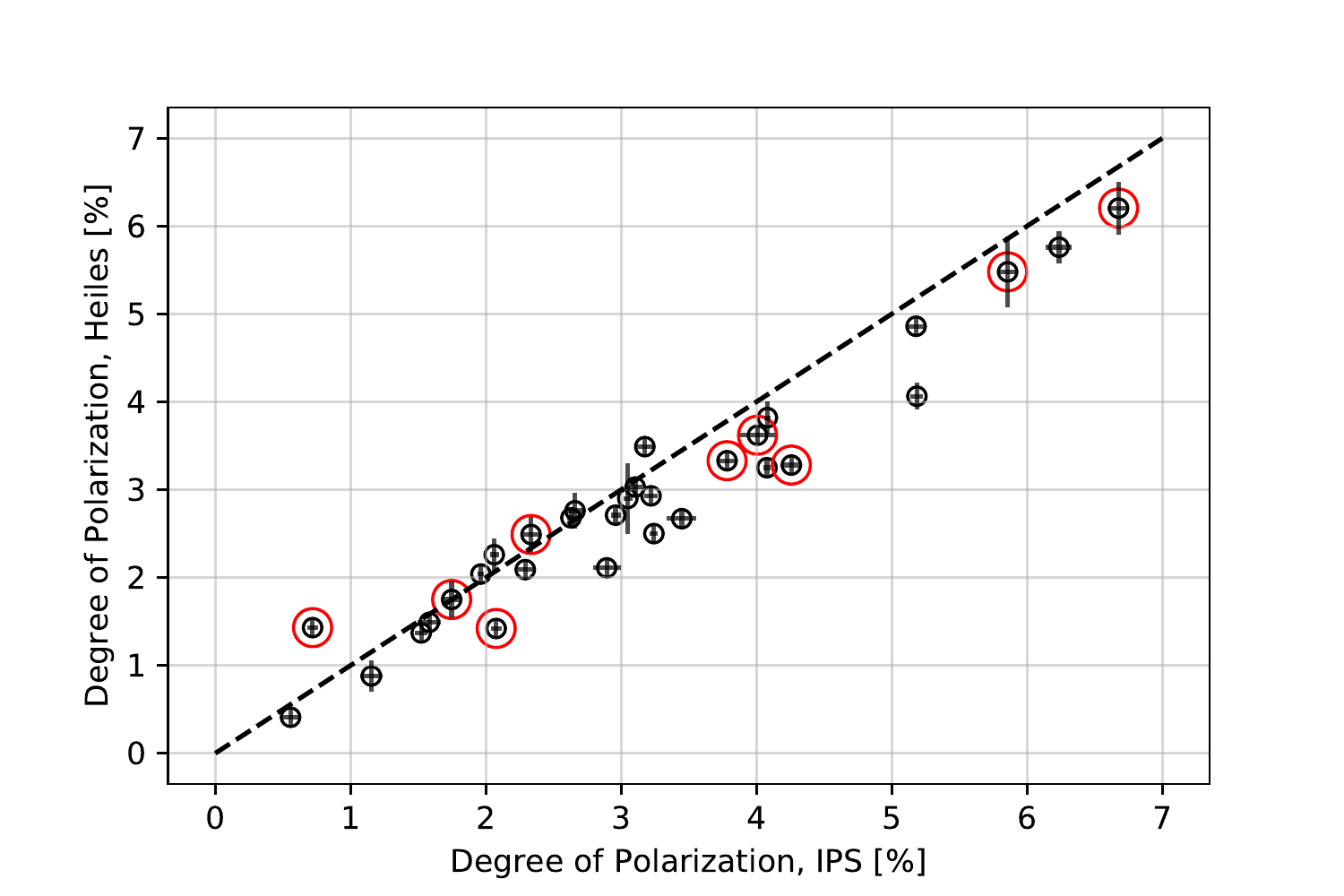}{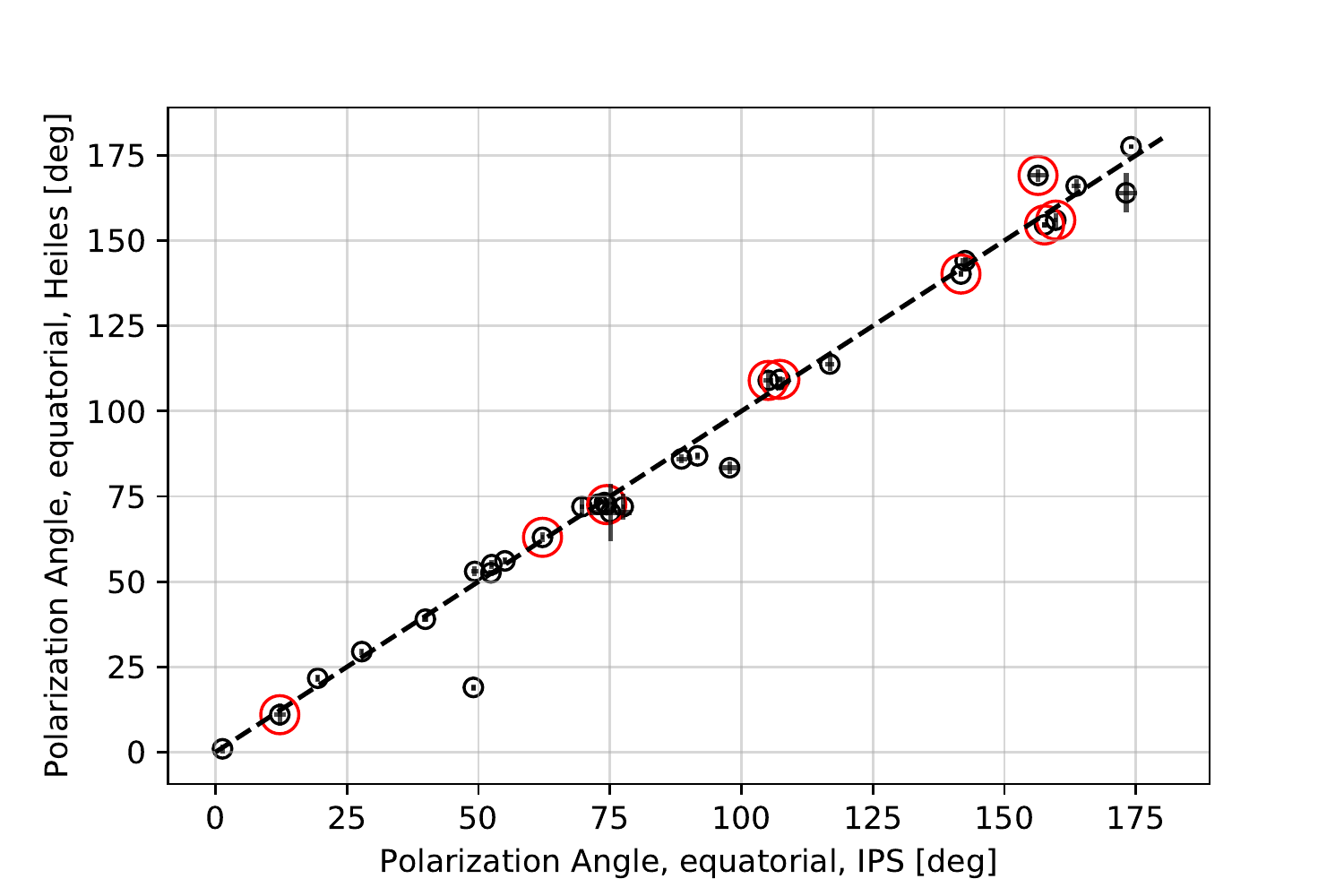}
\caption{Comparisons between the polarimetric measurements of \citet{heiles20009286} and IPS-GI. Red circles indicate variable stars. Left: degree of linear polarization. Right: angle of linear polarization. Dashed lines show the x=y diagonal. Error bars indicate the measurement uncertainties. The measurements of both data sets show strong agreement.}
\label{fig:heiles_comparison}
\end{figure*}

Although large databases of starlight polarization in the diffuse medium are rare, we compared our findings to other available measurements to verify their accuracy. To this end, we compared the IPS-GI \edit1{polarization values} with those from the \citet{heiles20009286} catalog. We used \textsc{Topcat} to match sources common to both catalogs. We found matches within 1.5 arcsec for \edit2{31} stars. The comparisons between the polarization degrees and polarization angles are shown in Fig. \ref{fig:heiles_comparison}.

Figure \ref{fig:heiles_comparison} shows good agreement between Heiles' catalog and the IPS-GI measurements. \edit2{Almost all} stars are O- and B-type, and \edit2{seven} are classified variable stars in the \emph{General Catalogue of Variable Stars} (GCSV, Version 5.1,\edit1{ \citealt{samus2017general}}). \edit2{In addition, one star is a Young Stellar Object (YSO) and another is a Wolf-Rayet star.} The variable stars \edit2{, as well as the YSO and WR star,} are marked red in Figure \ref{fig:heiles_comparison}. Many OB stars are known to exhibit polarimetric variability, see for example \citet{bjorkman1994spectropolarimetric}, which could partly explain the difference in measurements between the IPS and \citet{heiles20009286}. We note an apparent systemic deviation in seven stars, for which we found a \edit2{degree of} polarization higher than the Heiles measurement. One outlier stands out in the polarization angle comparison. The polarimetric measurements for this star from Heiles' catalog are taken from \citet{hiltner1956photometric}. This star, identified as BD-14 4922, has also been observed polarimetrically as part of the \citet{cikota2016linear} analysis of archival data of polarized standard stars. This study, which also includes V-band observations, finds a polarization angle for BD-14 4922 of $\theta =50 \pm 0.07\degr$. This is in agreement with our own polarimetry of this star, where we found a polarization angle of \edit2{$\theta = 49\pm 0.44 \degr$}. We therefore excluded this star from further comparison of the polarization angles with \citet{heiles20009286}. Furthermore, we also excluded all known variable stars from further comparison, see marked stars in Fig. \ref{fig:heiles_comparison}. \edit1{Variable stars are expected to also show polarimetric variability and are therefore not suitable for a comparison of this type.}
\par
\begin{figure*}[ht!]
\plottwo{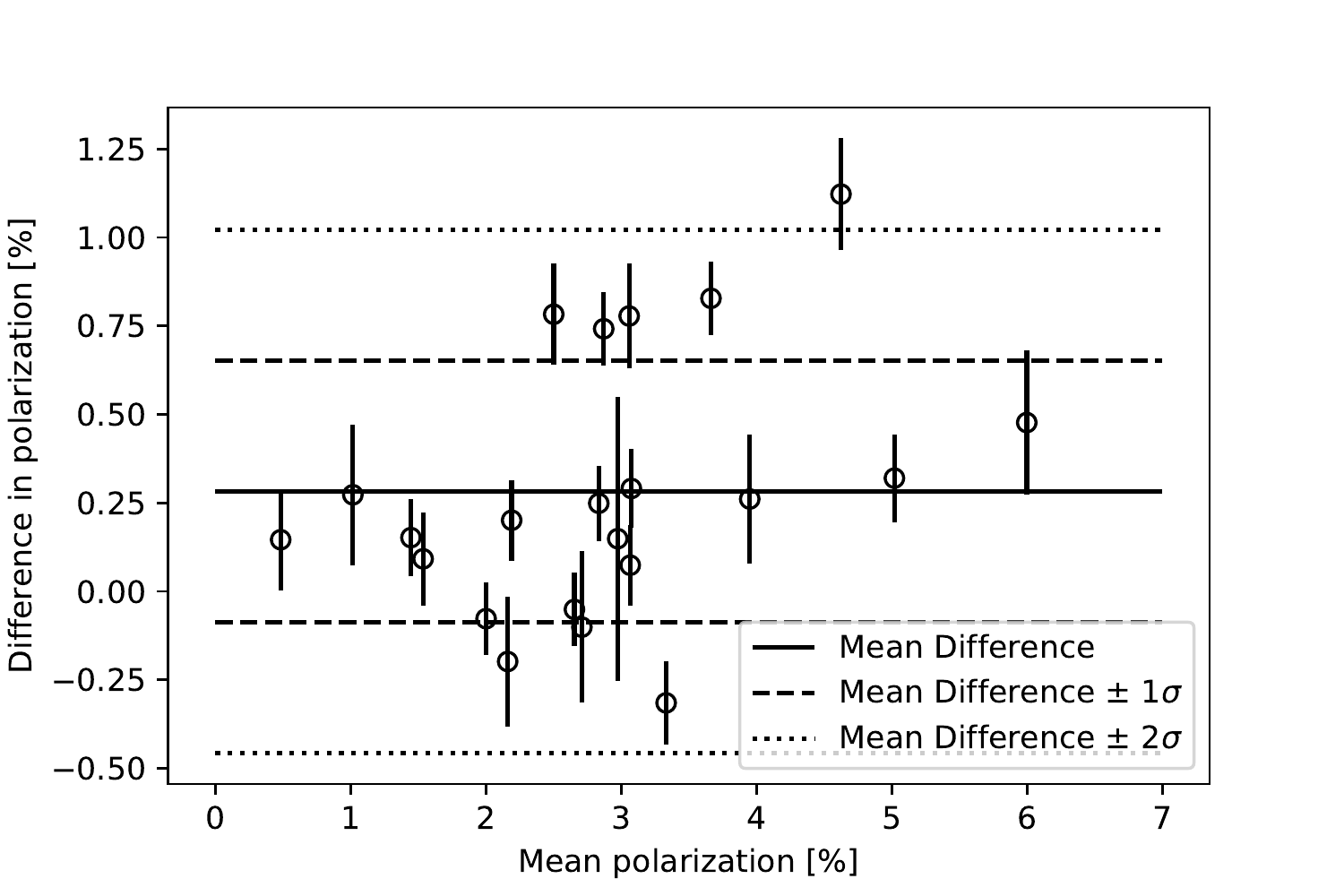}{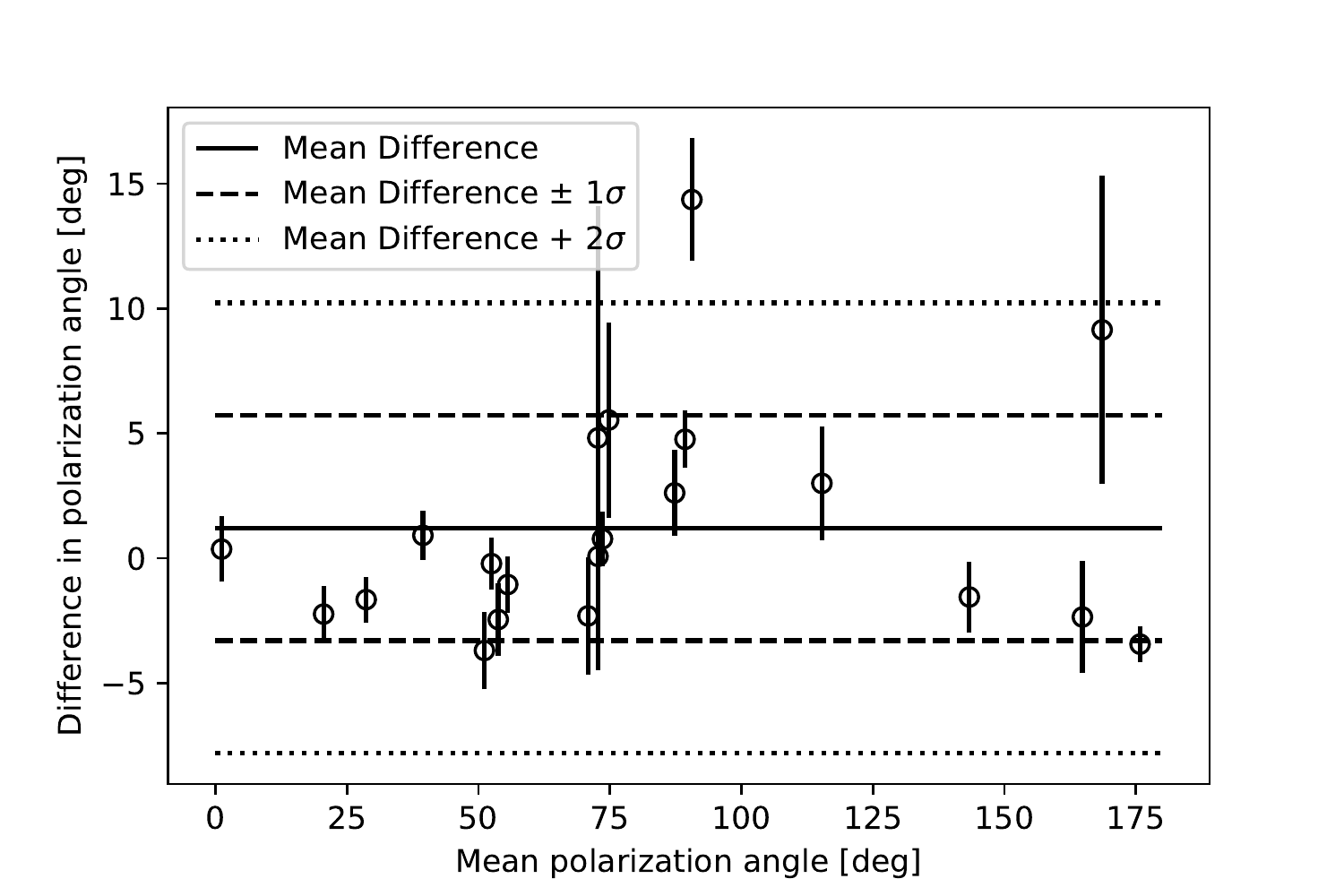}
\caption{Bland-Altman plots for the differences in measurements between Heiles and IPS-GI. Variable stars were removed from the comparison. Left: degree of linear polarization. Right: angle of linear polarization with BD-144922 removed. Solid lines indicate the mean difference between the two data sets. Dashed and dotted lines represent the 1$\sigma$ and 2$\sigma$ limits of agreement, respectively. Error bars indicate the errors of both measurements, added in quadrature.}
\label{fig:BA_comparison}
\end{figure*}

We calculated a modified reduced $\chi^{2}$ statistic using the following equation:

\begin{equation}
\chi^{2}_{red} = \frac{1}{N}\sum\frac{X_{dif, i}^{2}}{\sigma_{i}^{2}},
\end{equation}

\noindent where $N$ is the number of observations (\edit2{22} in the case of polarization degree, \edit2{21} for the polarization angle), $X_{dif}$ is the absolute difference between the observations, either the degree of polarization or the polarization angle, and $\sigma_{i}$ is the square root of the sum of squared error in both measurements. This lead to a $\chi^{2}_{red, P} = 12.2$ for the degree of polarization P, and a $\chi^{2}_{red, \theta} = 4.9$ for the polarization angle $\theta$. To visualize the differences between the measurements of the two catalogs, we created Bland-Altman plots \edit1{(see \citealt{bland1986statistical} for more details)} for both parameters. These are presented in Figure \ref{fig:BA_comparison}. Following the process outlined in \citet{bland1986statistical}, we defined a mean value, as well as limits of agreement out to 2$\sigma$ for both parameters. These plots show that for the degree of polarization P, \edit2{all stars have measured polarization (taking errors into account) that falls} within the 2$\sigma$ boundaries. For the polarization angle $\theta$, \edit2{only} one star falls outside this limit. \edit1{Because of the heterogeneous nature of the \citet{heiles20009286} compilation, as well as the varying quality and considerable age of some of the data therein, we are inclined to assign most differences between the catalogs to the uncertainties in the \citet{heiles20009286} agglomeration.}

\subsection{Applying quality filters \label{appfilters}}

\edit2{We have cross-matched the IPS data with two auxiliary catalogs: GAIA EDR3 \citep{gaia2021edr3} and Anders' et al. Starhorse-based catalog \citep{anders2022photo}. We wish to ensure that the analysis is applied only to the highest quality data. Therefore, we have applied six selection criteria that were defined based on the degree of linear polarization P and its error $\sigma_P$, various quality flags native to the auxiliary catalogs, and the cross-matching between the IPS-GI and GAIA EDR3 catalogs. These will be discussed in more detail below.}

\par
\edit2{
The first criterion, based on the degree of linear polarization P and its error \textsigma \textsubscript{P}, ensures that all sources under consideration have a high polarization signal-to-noise ratio: ${P}/\sigma_{P}>5$. Furthermore, using this filter, we avoid the statistically problematic region discussed in \citet{clarke1986statistical} and Section \ref{data:debias}. At high ${P}/\sigma_{P}$, the Rician bias produced by the positive nature of the polarization is mitigated  \citep[see for example][]{simmons1985point}. 
}

\par
\edit2{
Secondly, we retain only stars with small absolute polarimetric errors, i.e. $\sigma_P\le0.8\%$. This filter removes stars with spurious polarimetric measurements.
}
\par
\edit2{
Thirdly, we filter for spurious GAIA results using the \emph{fidelity} parameter, see also \citet[][]{rybizki2022classifier}. Applying $\textrm{fidelity}>0.5$ removes stars with poor astrometric solutions.}
\par
\edit2{
$\textrm{SH}_{outflag}$ is an \citet{anders2022photo} native-parameter and filtering for outflag=0000 is recommended by the authors. This removes spurious results caused by unreliable extinctions, overly large distances and poor \textsc{StarHorse} convergence. 
}
\par
\edit2{
The fifth filter is based on recommendations of \citet[][]{riello2021photometric}, see Sections 6 and 9.4 therein.  The color excess factor described in \citet[][]{riello2021photometric} can be used to check for inconsistencies in the various photometries, allowing for the exclusion of spurious sources.
}
\par
\edit2{
Finally, we remove any sources for which there is more than once possible match in the GAIA catalog. Despite not only cross-matching based on spatial position but also magnitude, some stars have multiple potential matches. These stars are removed from further consideration to ensure only certain matches are used in the analysis.}
\edit2{
In sum, the selection criteria are as follows:
}
\begin{enumerate}
    \item $p/\sigma_{p} > 5$, which ensures only high-quality polarimetric observations are taken into consideration,
    \item $dp < 0.8\%$, which removes sources with spurious polarimetric errors,
    \item $fidelity > 0.5$, which removes spurious results from GAIA fitting routines, see \citet{rybizki2022classifier},
    \item $sh_{outflag} = 0$, which removes spurious results from \citet{anders2022photo}, 
    \item $C^{*}/\sigma_{C} < 5$, which exclude sources with spurious color excess, see also \citet[][eq. 18 therein.]{riello2021photometric}
    \item Only high-certainty matches, which removes results for which the cross-matching between the IPS and GAIA EDR3 catalogs was uncertain.
\end{enumerate}

\edit2{
Applying all six criteria, we retained a subsample of 10516 sources to study in more detail. On average, each field will contain $\sim275$ filtered stars.}

\section{Statistical properties of the data}
\label{sec:statprop}
\subsection{Low polarization sources \label{data:lowpol}}
The filtered subsample shows a lack of low polarization sources. This appears to be caused by the  ${P}/\sigma_{P} > 5$ filter. This introduced a bias towards stars with higher polarization, inadvertently filtering out low polarization sources. However, although stringent, this filter ensures that all analysis is applied only to the highest quality observations. In addition, it removes the need for debiasing as explained in section \ref{data:debias}.

\subsection{High polarization sources \label{data:highpol}}

\edit2{After filtering, a single star remains that exhibits a very high degree of polarization, i.e. $P>10\%$. This high polarization measurements may be indicative of intrinsic polarization, but may also be a result of, for example, a favorable geometry of the magnetic field. This has been shown to lead to an increase in the polarized signal from stars, see for example \citet{panopoulou2019extreme}. Further investigation of this individual star, as well as other stars that show extraordinary polarimetry, is beyond the scope of this paper.}

\subsection{Spatial distributions \label{data:spat}}

Assuming the starlight to be initially unpolarized, all measured polarization is a direct result of starlight passing through the ISM. As the polarization is a function of the location where it is produced, which can be defined by Galactic longitude \emph{l}, Galactic latitude \emph{b} and distance \emph{d}, it is important to study the spatial distribution of the polarization.
\par
\begin{figure}[ht!]
\plotone{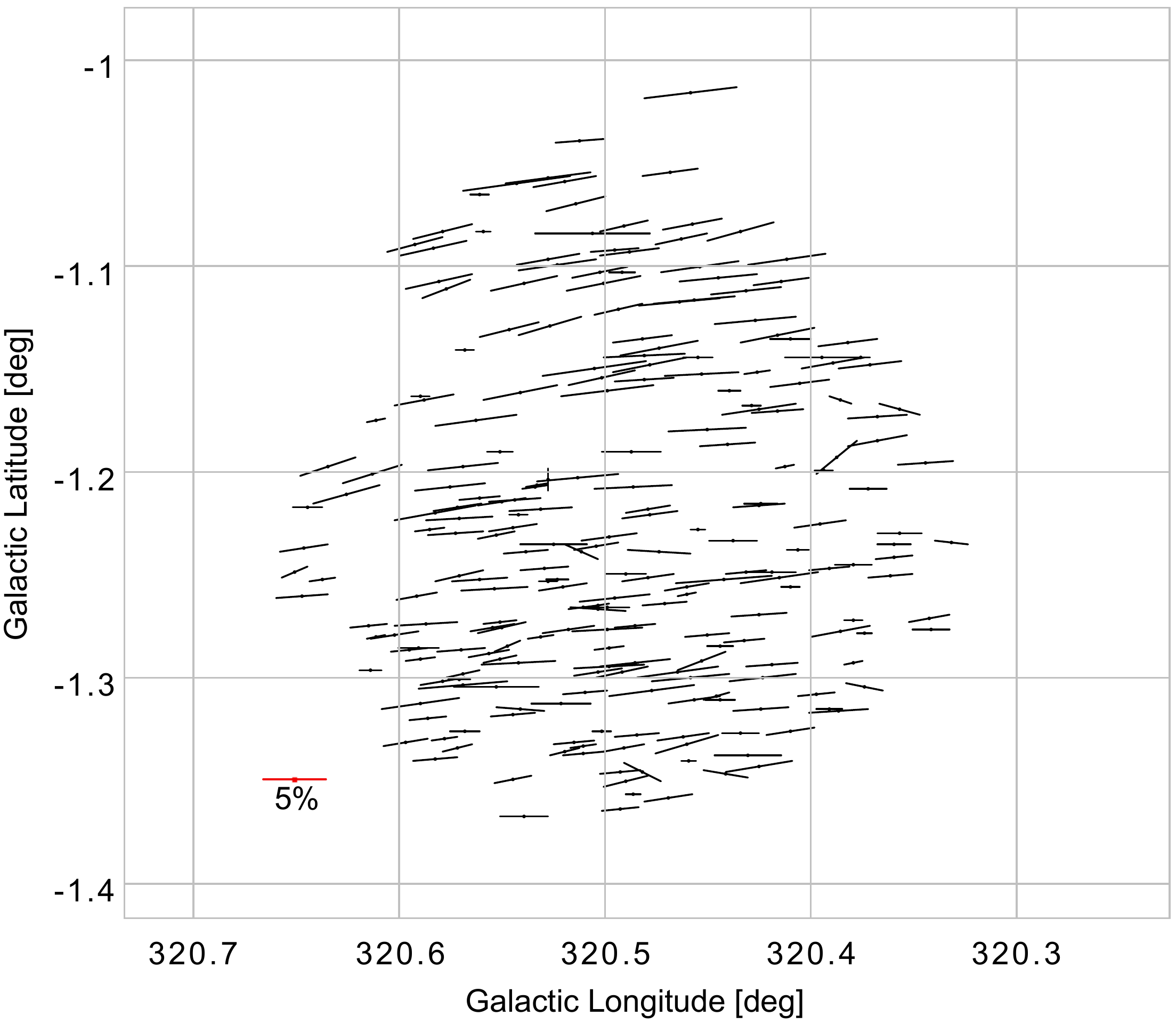}
\caption{Example field C47 containing \edit2{240} stars, located at l=320.5\degr, b=-1.2\degr. The vectors represent the polarization angle and degree. Red example vector shows a polarization degree of $5\%$ and $\theta_{gal}=90\degr$. See Fig. \ref{fig:c47_pol_distribs} for more details.}
\label{fig:campo_47_example}
\end{figure}
\begin{figure*}[ht!]
\plotone{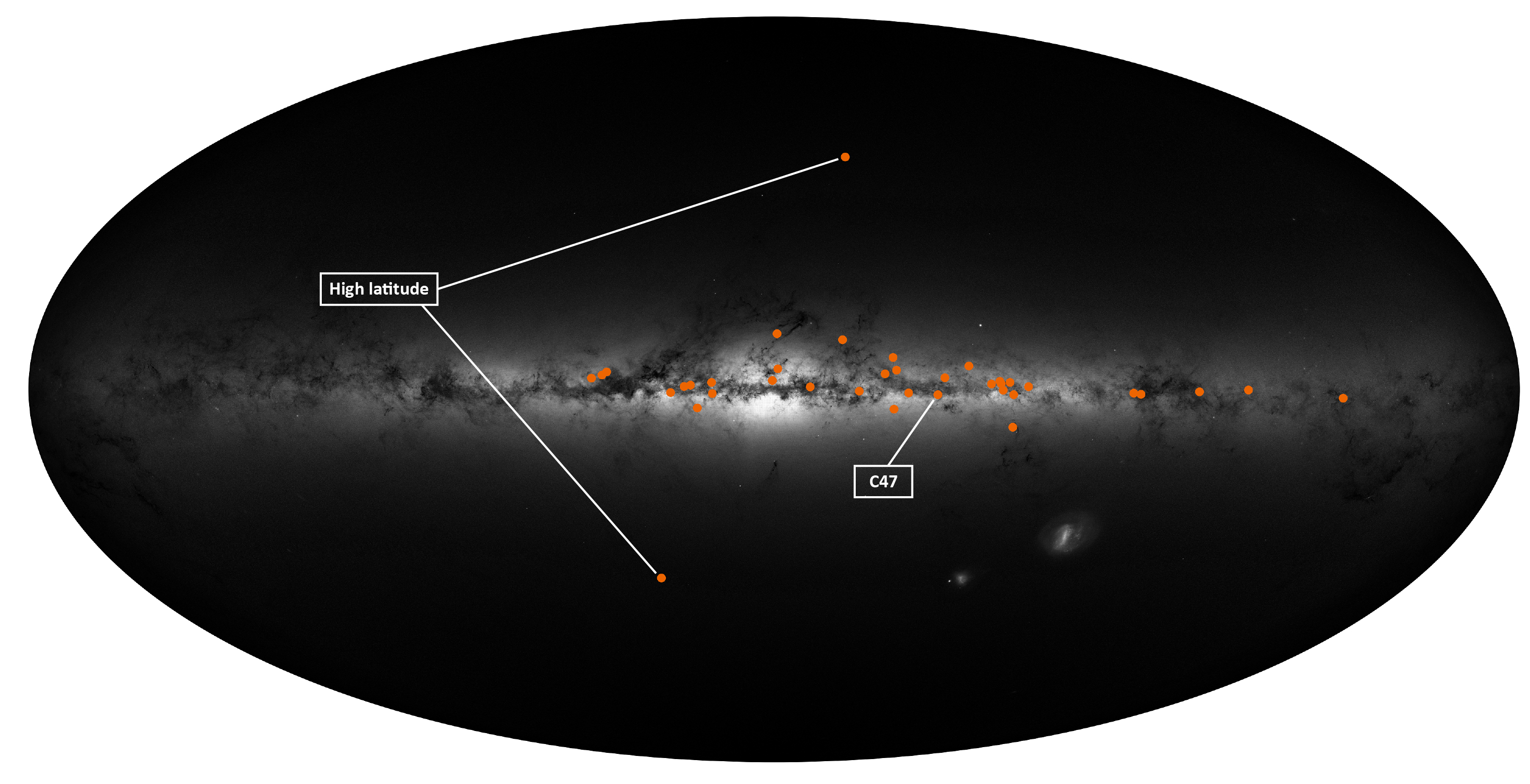}
\caption{Positions of lines of sight in IPS-GI projected on the Milky Way. Each marking represents an observed field of stars. Longitude increases towards the right. Labels indicate the two high Galactic latitude fields and example field C47. Background image adapted from ESA/Gaia/DPAC.}
\label{fig:los_mw}
\end{figure*}

The IPS-GI observations are focused on certain lines of sight in the diffuse ISM, which leads to a very inhomogeneous source distribution across the sky, see Fig \ref{fig:los_mw}. Figure \ref{fig:campo_47_example} shows an example of a typical ordered field (C47, cf Table \ref{tab:perfield}) located at $l=320.5\degr$, $b=-1.2\degr$. After applying quality filters (see Section \ref{appfilters}), this field contains \edit2{240} stars. Each star's polarization vector was plotted-- its orientation is equal to the Galactic polarization angle and its length depends on the degree of polarization. The uniformity of the polarization vectors is typical for this data set\edit1{ (see middle panel of Fig. \ref{fig:c47_pol_distribs}),} and is indicative of a dominant large-scale magnetic field even over the full range of distances, see Figs. \ref{fig:c47_pol_distribs}, middle and right.
\par

\par
\begin{figure*}[ht!]
\gridline{\fig{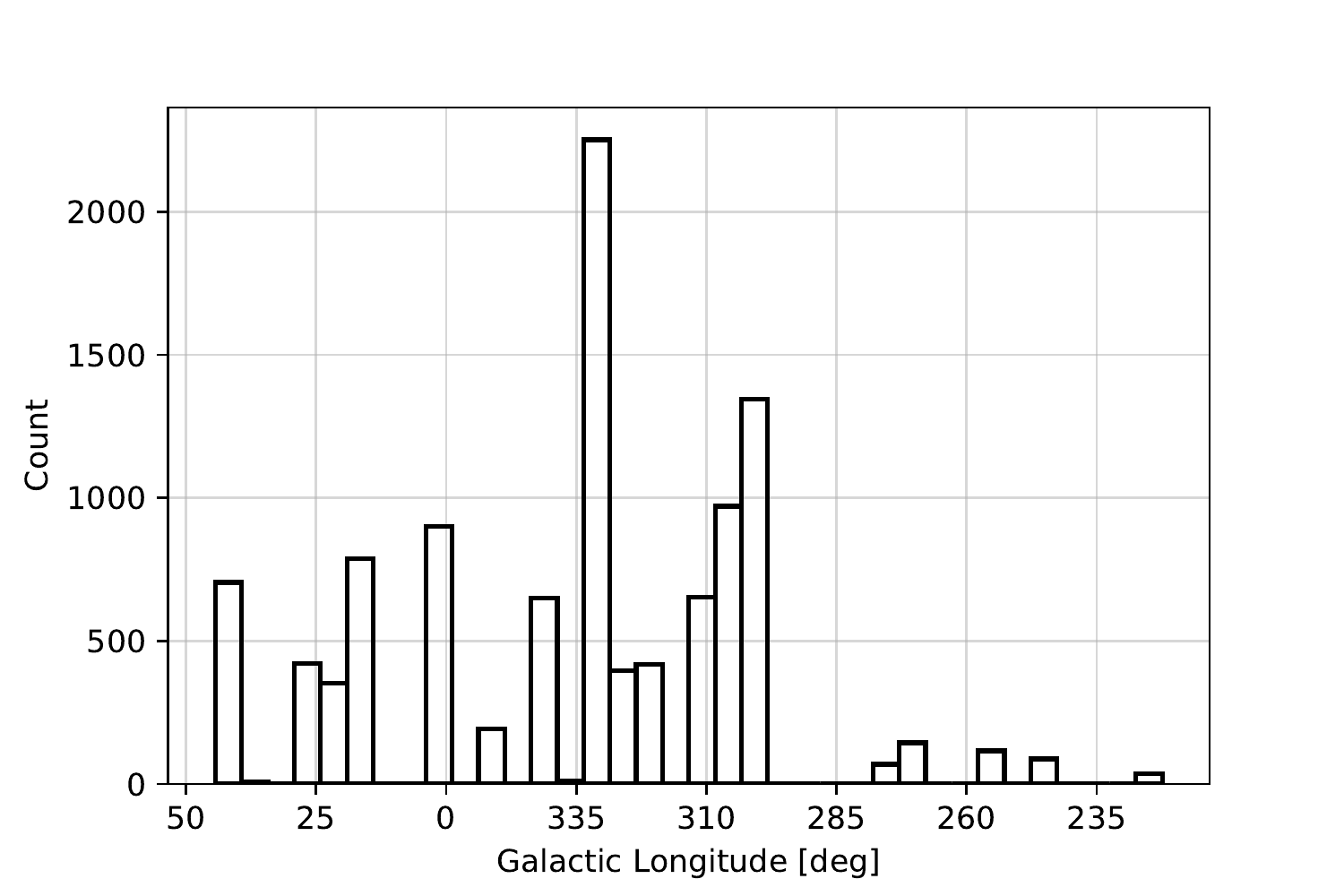}{0.3\textwidth}{(a)}
          \fig{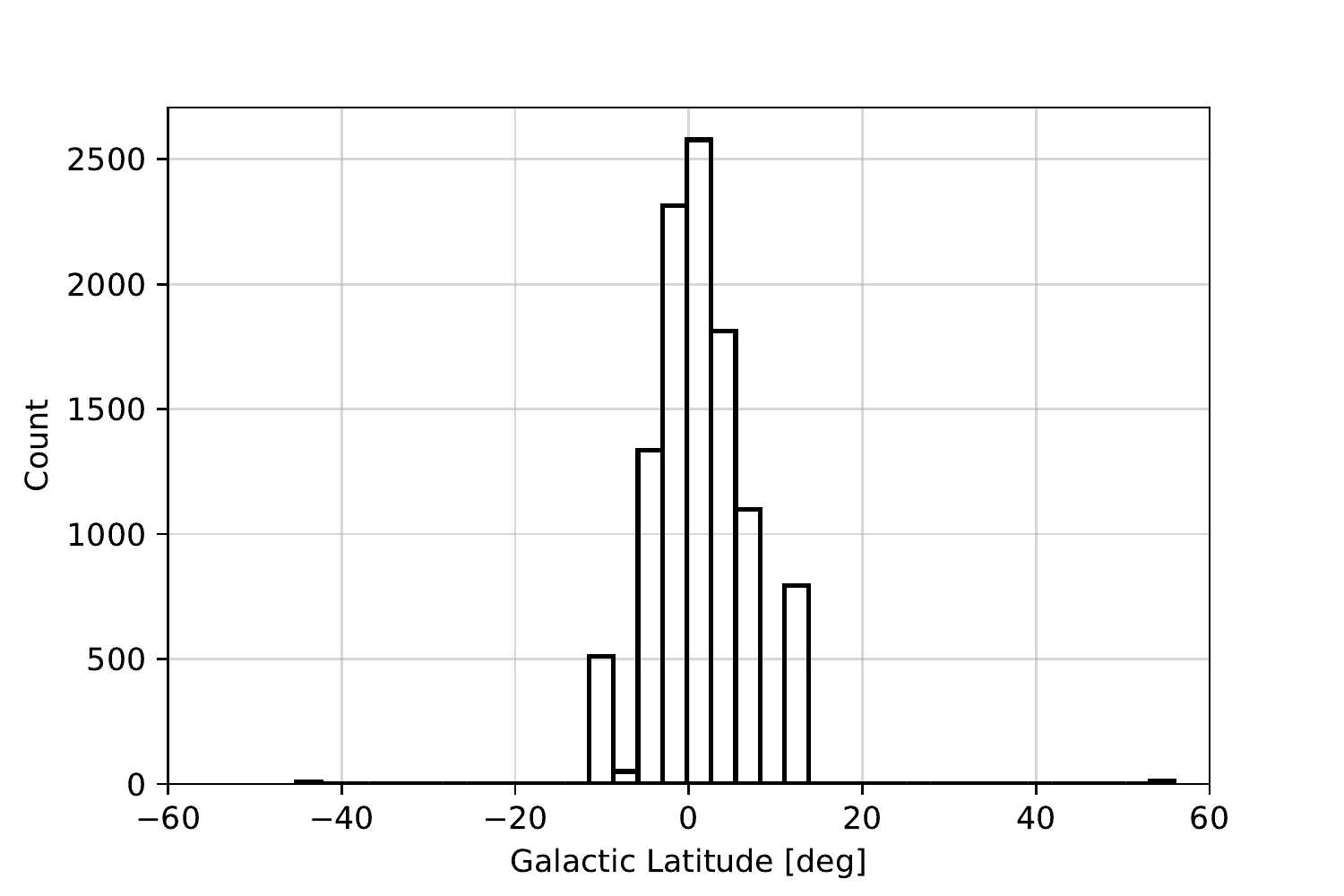}{0.3\textwidth}{(b)}
          \fig{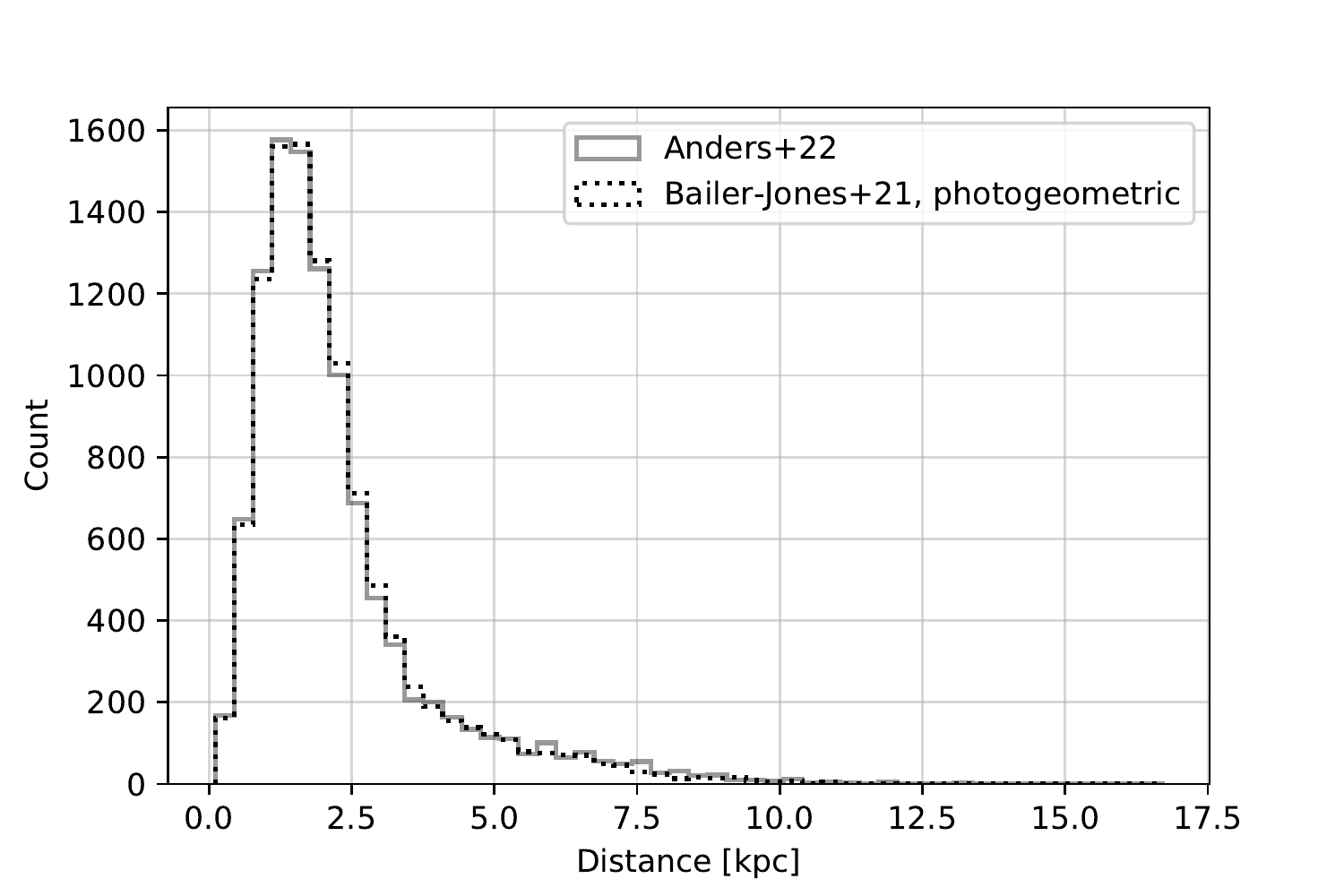}{0.3\textwidth}{(c)}
          }
\caption{Distributions of spatial parameters for all filtered IPS-GI data. Left: Galactic longitude \emph{l} in degrees. Middle: Galactic latitude \emph{b} in degrees. \edit2{Right: distance \emph{d} in kpc. Solid line data is taken from \citet{anders2022photo}, the dotted line represents (photogeometric) distances from \citet[]{bailer2021estimating}}.}
\label{fig:spatial_distribs}
\end{figure*}

Figure \ref{fig:los_mw} shows the locations of all observed fields within the Milky Way. In addition, the distributions for Galactic longitude, Galactic latitude and distance in Fig. \ref{fig:spatial_distribs} show that the observed fields lie within a longitude range of $220\degr  < l <  50\degr$ and that most fields are within 15\degr\ latitude from the Galactic plane. Two sparsely populated fields are positioned further above and below the Galactic plane, see annotations in Fig. \ref{fig:los_mw}. We also indicated the location of example field C47 (see Fig. \ref{fig:campo_47_example}). Finally, most stars are located nearby (around $d \sim 2\textrm{ kpc}$), with \edit2{significant numbers out to $d = 6\textrm{ kpc}$, see Fig. \ref{fig:spatial_distribs}, right, with certain stars at even higher distances}.

\subsection{Photometric and polarimetric distributions \label{data:photpol}}

\begin{figure*}[ht!]
\gridline{\fig{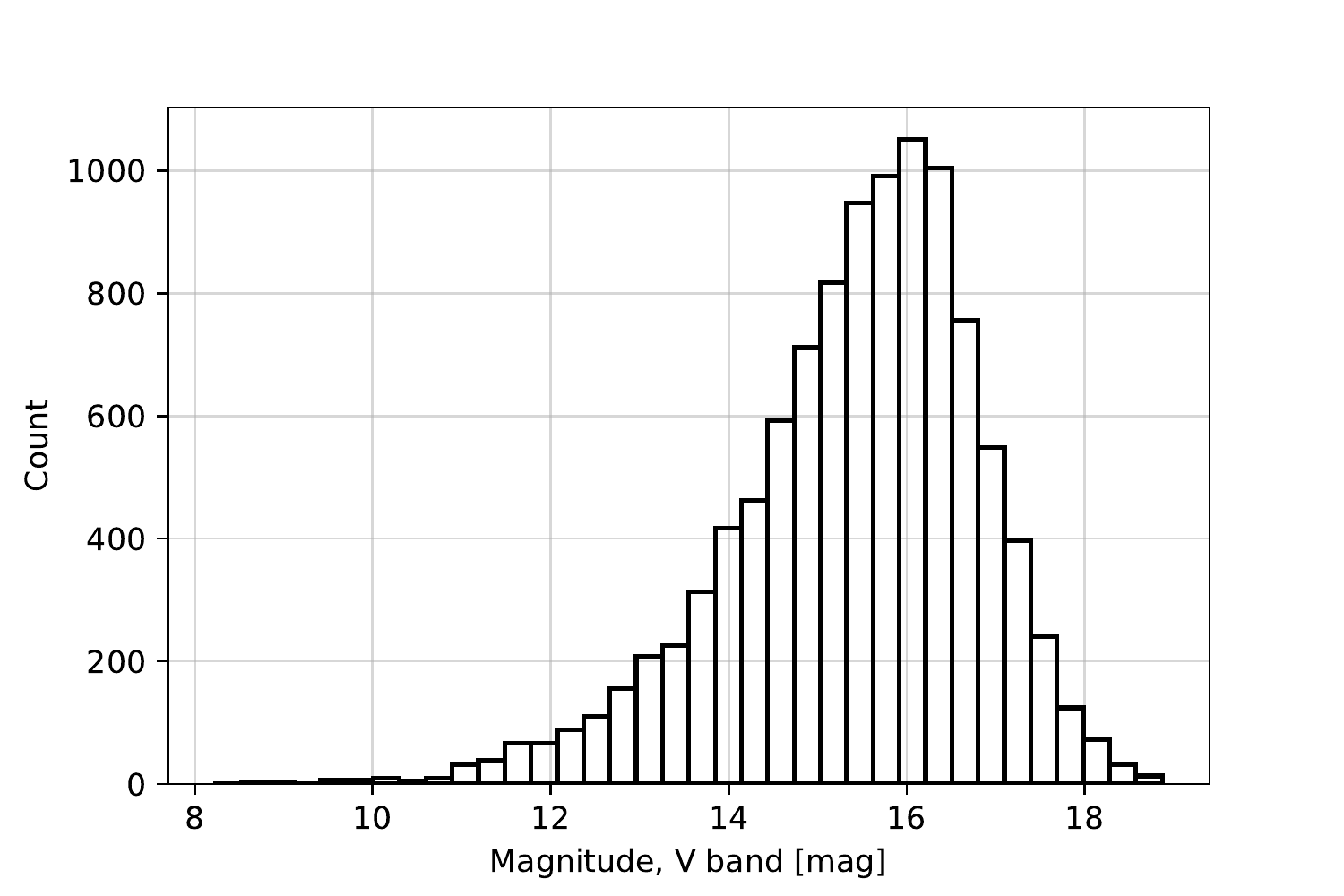}{0.3\textwidth}{(a)}
          \fig{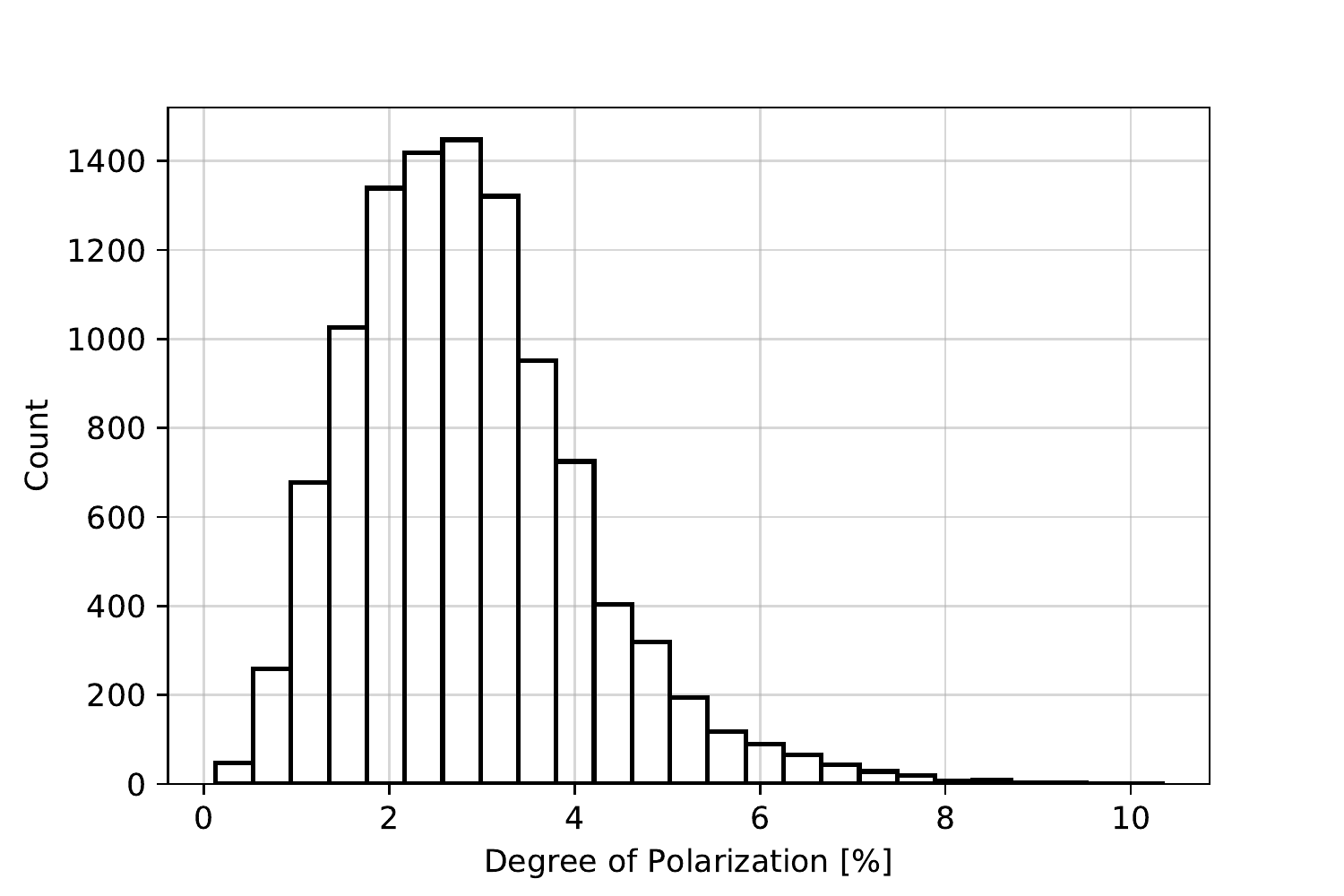}{0.3\textwidth}{(b)}
          \fig{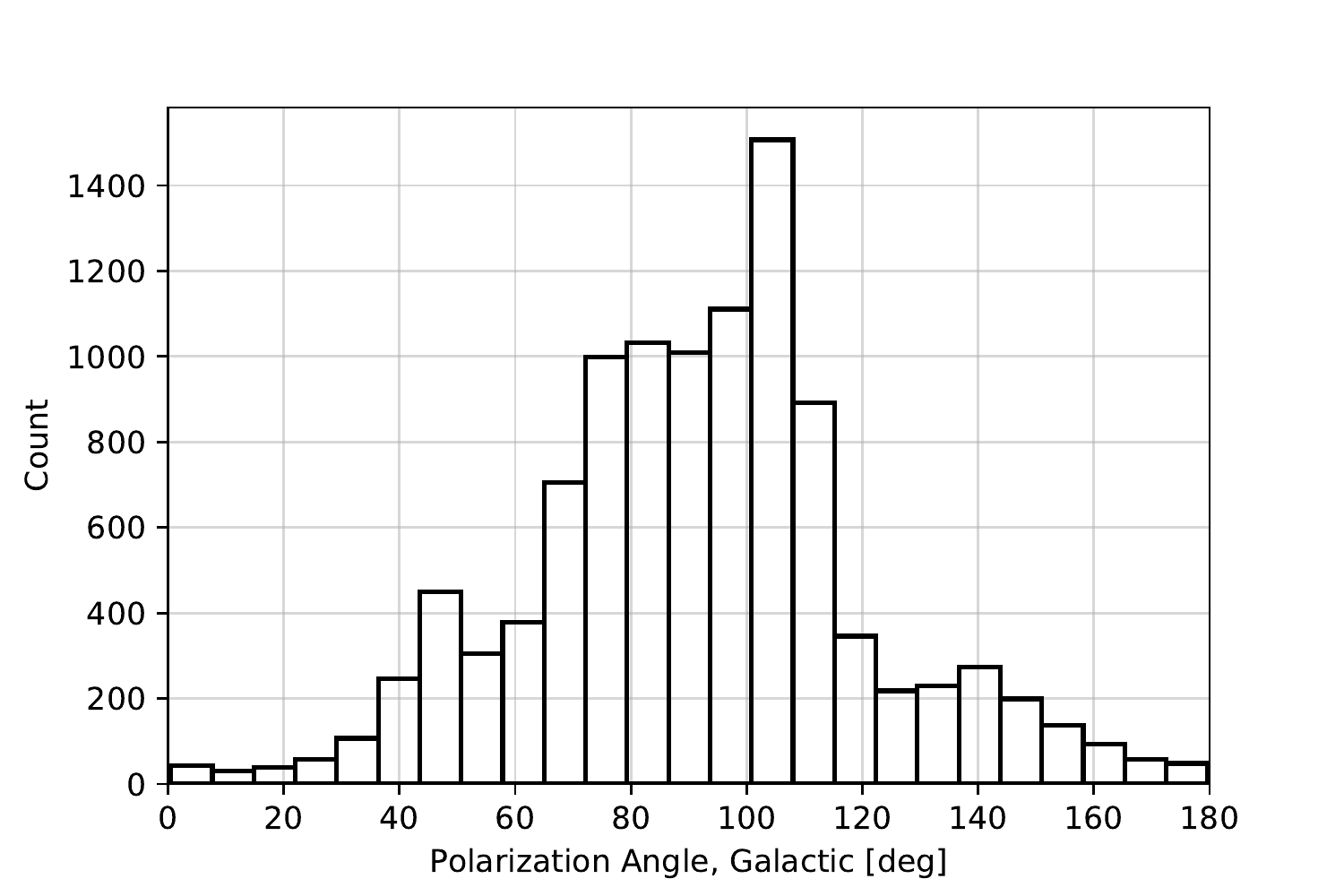}{0.3\textwidth}{(c)}
          }
\caption{Distributions of photometric and polarimetric parameters for all filtered IPS-GI data. Left: V-band magnitude. Middle: degree of polarization P as percentage. Right: angle of polarization $\theta$ in degrees in the Galactic coordinate system. $\theta_{Gal} = 90 \degr$ is oriented along the Galactic plane.}
\label{fig:photpol_distribs}
\end{figure*}

Next, we looked at the distributions of the V-band magnitude and the polarimetric parameters P and $\theta$. Figure \ref{fig:photpol_distribs}, left, shows the distribution of the magnitudes for the filtered survey data: IPS-GI covers a range of magnitudes roughly from 8 to 19. The majority of stars have V \textless 15. We again note that this survey is not complete in magnitude due to the different integration times of different fields.
\par
Figures \ref{fig:photpol_distribs}, middle and right, show the distributions of the degree of polarization P and the angle of polarization $\theta$, respectively. The degree of polarization peaks around 3\%, with a notable absence of very low polarization sources (see Section \ref{data:lowpol}). Furthermore, some stars show very high (P \textgreater 10\%) polarization (see Section \ref{data:highpol}). The distribution of polarization angles peaks around 90\degr, which is expected for a line-of-sight-averaged magnetic field that is parallel to the Galactic plane. This is in agreement with observations based on radio polarimetry (see for example \citet{haverkorn2015milkyway}) as well as observations of external spiral galaxies \citep{beck2015spiral}. The distribution also shows a second peak around $\theta = 40\degr$ and a somewhat flattened structure around $\theta = 140\degr$. These features will be discussed in more detail in Section \ref{avgperfield}.

\par
\begin{figure*}[ht!]
\gridline{\fig{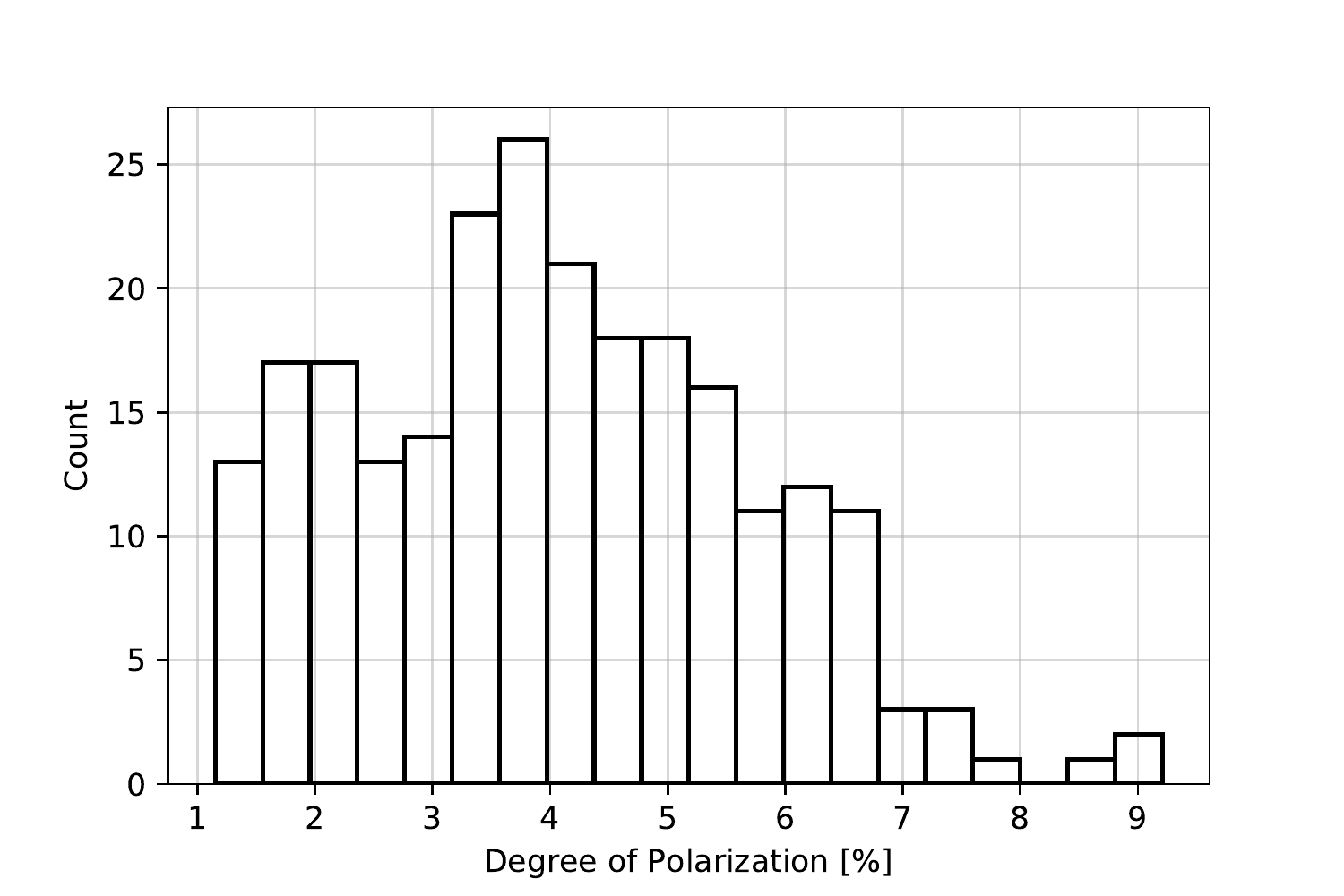}{0.3\textwidth}{(a)}
          \fig{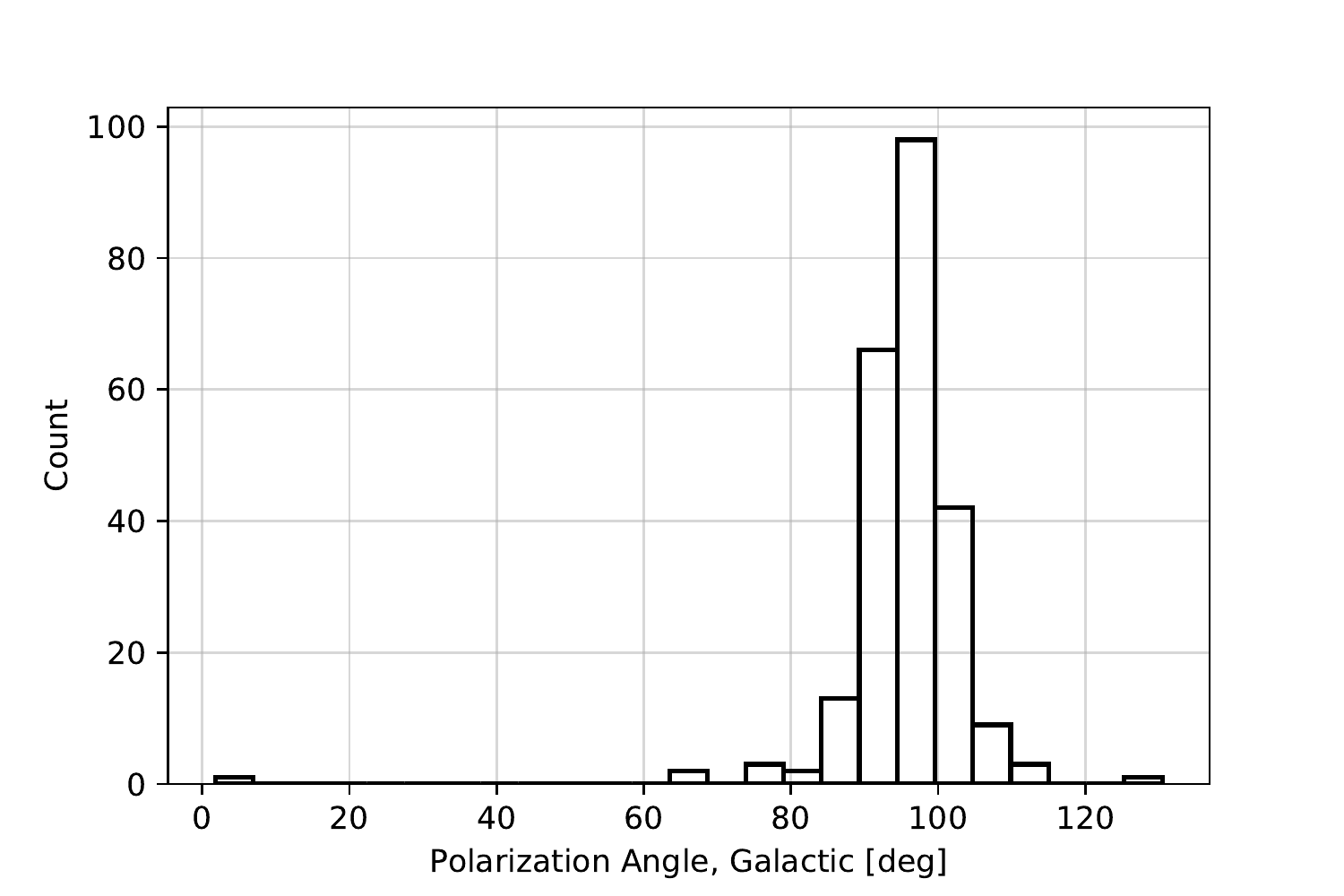}{0.3\textwidth}{(b)}
          \fig{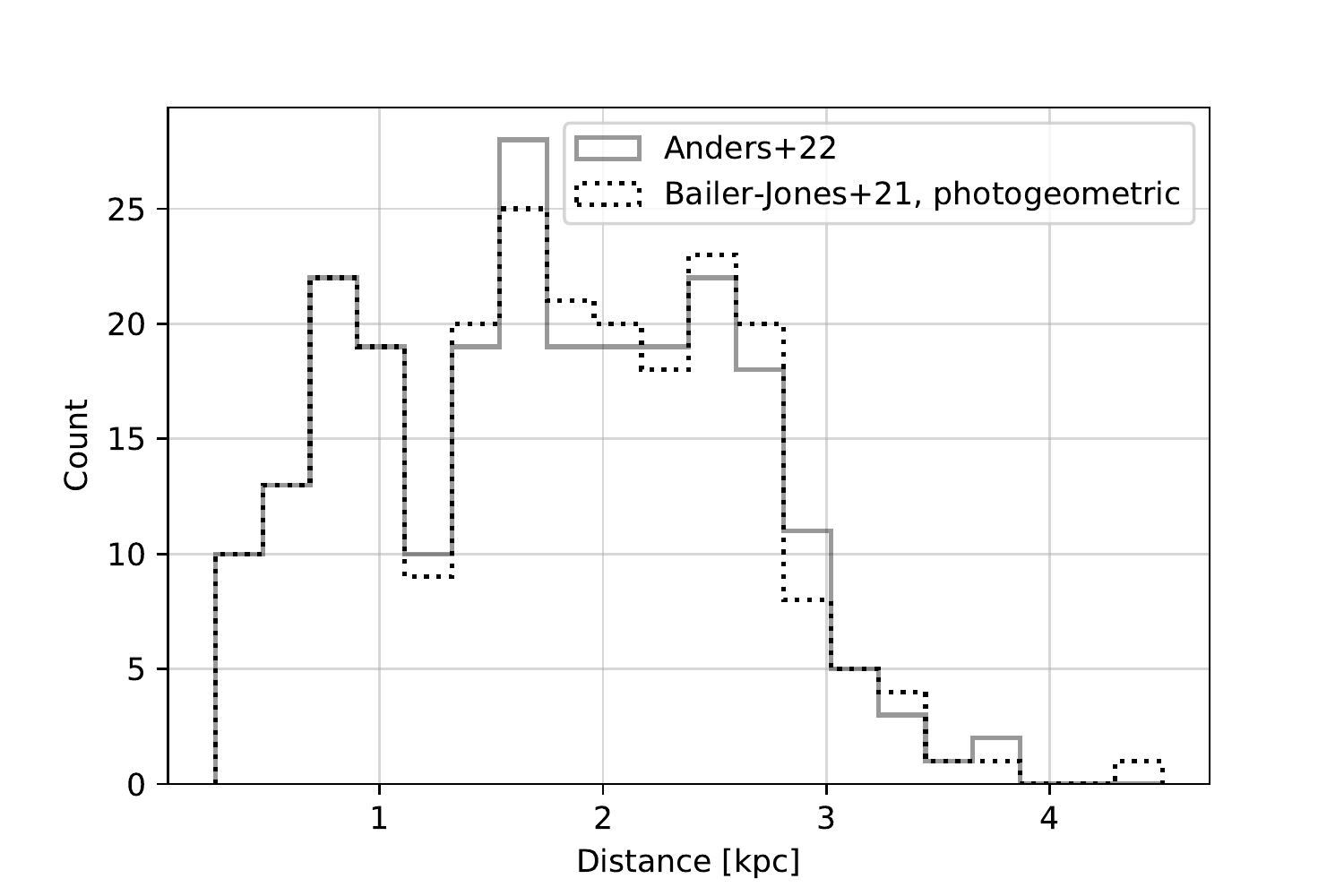}{0.3\textwidth}{(c)}
          }
\caption{Distributions of polarimetric parameters and distances for example field C47. Left: degree of polarization \emph{P} as percentage. Middle: angle of polarization \emph{\straighttheta} in degrees in the Galactic coordinate system. \edit2{Right: distance \emph{d} in kpc. Solid line data is taken from \citet{anders2022photo}, the dotted line represents (photogeometric) distances from \citet[]{bailer2021estimating}.}}
\label{fig:c47_pol_distribs}
\end{figure*}

Returning to our example field C47, we took a closer look at the distributions of the degree of polarization and the polarization angle. Histograms showing the distributions of those parameters are presented in Figure \ref{fig:c47_pol_distribs}. \edit2{We include a comparison of different distances for this field in Fig. \ref{fig:c47_pol_distribs} (right).} These figures quantify the structure seen in Figure \ref{fig:campo_47_example}. The degrees of polarization vary over a relatively wide range, but the polarization angle is highly peaked around an average $\theta_{Gal} = 96\degr$.

\subsection{Averages per field \label{avgperfield}}
Table \ref{tab:perfield} describes the main properties of the IPS-GI filtered subsample. The weighted averages of the polarimetric parameters P and \straighttheta were calculated following conventions from \citet{pereyra2007polarimetry}.
From this table, we found that some fields have an average orientation that deviates greatly from $\theta_{Gal} \sim 90\degr$, a value expected for an average magnetic field direction in the plane of the Milky Way. Some of these fields account for features mentioned above, e.g. the histogram of polarization angles in Figure \ref{fig:photpol_distribs}, right. For example, field C4 has an average Galactic polarization angle of $\theta_{Gal} = 45\degr$. With 455 stars, this field contributes significantly to the shape of Figure \ref{fig:photpol_distribs}, right. Furthermore, fields C7, C50 and C57 have average polarization angles of $\theta_{Gal} = 155\degr$, $\theta_{Gal} = 132\degr$ and $\theta_{Gal} = 142\degr$, respectively. These fields explain the apparent flattening of the histogram in Figure \ref{fig:photpol_distribs}, right. 
\par
\begin{figure*}[ht!]
\plotone{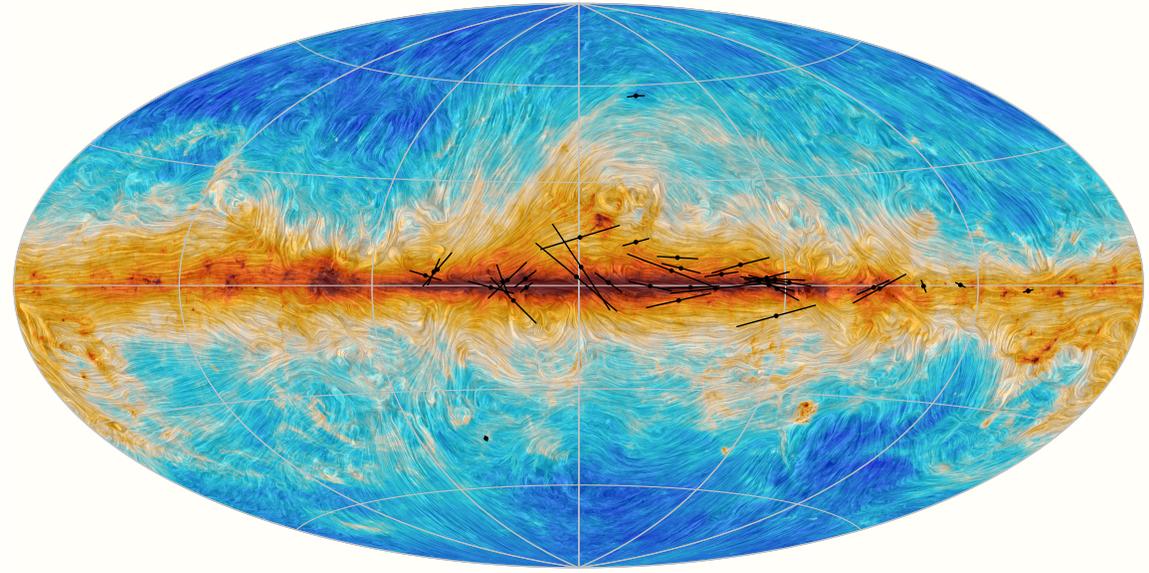}
\caption{Sky plot with vectors of weighted average degree of polarization (length) and weighted average polarization angle (orientation) per field. Background image copyright ESA/Planck collaboration, credit Marc-Antoine Miville-Deschênes. Longitude increases toward the right.}
\label{fig:skyplot}
\end{figure*}
\begin{figure}[ht!]
\plotone{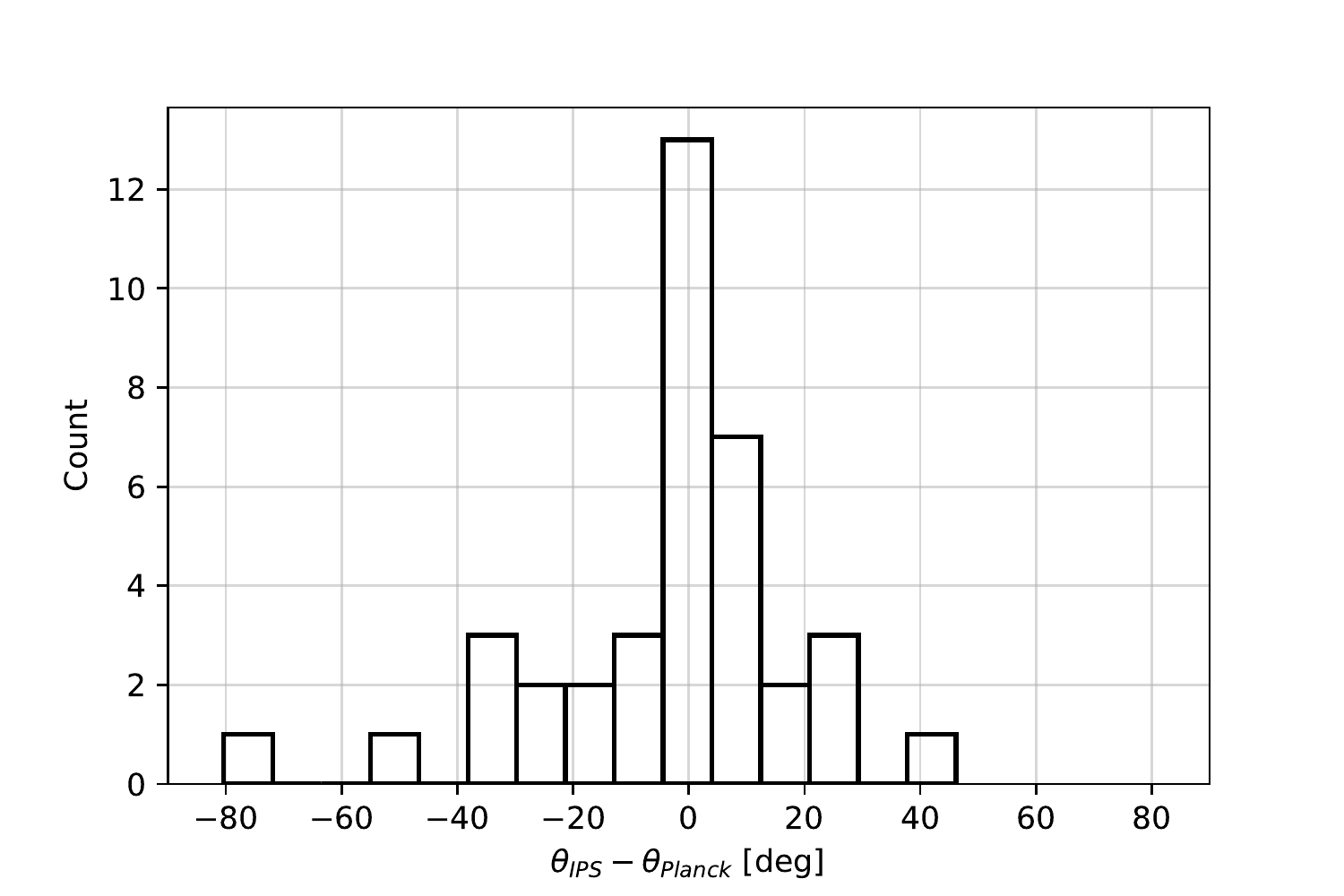}
\caption{\edit2{Histogram indicating the differences between average polarization angles per field from IPS ($\theta_{IPS}$) and Planck ($\theta_{Planck}$) in the same region.}}
\label{fig:plank_comp}
\end{figure}

Table \ref{tab:perfield} has also been summarized in Figure \ref{fig:skyplot}, which \edit1{qualitatively} shows the average polarization vectors per field on a sky plot. The background shows the Galactic magnetic field as detected by Planck. The optical polarization vectors are mostly in agreement with an ordered large-scale magnetic field parallel to the Galactic plane as can be seen in the background. \edit2{The differences between polarization angles as observed by IPS and Planck are shown in Figure \ref{fig:plank_comp}, indicating agreement typically within $\Delta\theta\sim10\degr$.} Any turbulent components along the line of sight at different distances are cancelled out or diminished. This can only be concluded based on starlight polarization data in conjunction with distance measurements, as Planck and similar sub-mm surveys show the result of an integration along an entire line of sight.

\startlongtable
\begin{deluxetable*}{cllllllllllll}
\tablenum{1}
\tablecaption{Mean values per IPS-GI field\label{tab:perfield}}
\tablewidth{0pt}
\tablehead{
\colhead{Field ID \tablenotemark{a}} & \colhead{l} & \colhead{b} & \colhead{P} & \colhead{dP} &  \colhead{$\theta_{Gal}$} & \colhead{d$\theta$} & \colhead{$\sigma\theta_{Gal}$} & \colhead{V} & \colhead{$A_{V}$} & \colhead{E(B-V)} & \colhead{Dist} & \colhead{N stars} \\
\colhead{} & \colhead{(deg)} & \colhead{(deg)} & \colhead{(\%)} & \colhead{(\%)} & \colhead{(deg)} & \colhead{(deg)} & \colhead{(deg)} & \colhead{(mag)} & \colhead{(mag)} & \colhead{(mag)} & \colhead{(kpc)} & \colhead{} \\ \hline
}
\decimalcolnumbers
\startdata
C0 & 327.57 & -0.83 & 2.44 & 0.15 & 92.08 & 1.79 & 12.81 & 15.22 & 1.79 & 0.58 & 1.49 & 397 \\  
C1 & 330.42 & 4.59 & 1.62 & 0.13 & 76.03 & 2.23 & 12.17 & 15.61 & 1.75 & 0.56 & 2.39 & 545 \\  
C2 & 359.34 & 13.47 & 3.14 & 0.16 & 107.62 & 1.49 & 9.41 & 16.51 & 1.11 & 0.36 & 3.52 & 594 \\  
C3 & 339.49 & -0.42 & 1.72 & 0.11 & 77.36 & 1.88 & 13.53 & 14.98 & 1.94 & 0.63 & 1.59 & 448 \\  
C4 & 18.63 & -4.46 & 2.46 & 0.18 & 45.67 & 2.12 & 8.64 & 15.14 & 1.54 & 0.5 & 2.91 & 455 \\  
C5 & 25.06 & -0.74 & 1.68 & 0.12 & 78.9 & 1.96 & 14.55 & 15.5 & 2.23 & 0.72 & 1.97 & 421 \\  
C6 & 298.61 & 0.64 & 1.66 & 0.14 & 79.44 & 2.45 & 15.32 & 14.25 & 1.43 & 0.46 & 1.85 & 210 \\  
C7 & 41.68 & 3.39 & 1.18 & 0.13 & 155.16 & 3.15 & 15.52 & 15.74 & 2.3 & 0.74 & 2.62 & 326 \\  
C11 & 307.53 & 1.33 & 1.5 & 0.1 & 76.0 & 1.85 & 14.32 & 14.58 & 1.32 & 0.43 & 1.87 & 372 \\  
C12 & 331.04 & -4.7 & 2.6 & 0.13 & 103.04 & 1.41 & 8.29 & 15.28 & 1.33 & 0.43 & 3.0 & 881 \\  
C13 & 333.24 & 3.75 & 3.81 & 0.25 & 69.99 & 1.91 & 6.11 & 15.84 & 2.53 & 0.82 & 1.73 & 287 \\  
C14 & 312.87 & 5.48 & 2.23 & 0.12 & 104.62 & 1.53 & 11.61 & 15.79 & 1.24 & 0.4 & 3.04 & 653 \\  
C15 & 304.71 & -0.17 & 2.16 & 0.11 & 77.46 & 1.46 & 13.67 & 15.8 & 2.48 & 0.8 & 1.91 & 461 \\  
C16 & 301.97 & -8.77 & 3.13 & 0.14 & 104.99 & 1.3 & 7.43 & 15.44 & 1.03 & 0.33 & 2.4 & 561 \\  
C30 & 222.54 & -1.63 & 0.33 & 0.09 & 84.32 & 7.9 & 35.57 & 13.96 & 1.44 & 0.47 & 1.7 & 37 \\  
C34 & 34.81 & -45.02 & 0.27 & 0.16 & 11.12 & 16.78 & 35.26 & 15.38 & 0.17 & 0.05 & 1.15 & 9 \\  
C35 & 271.45 & -1.07 & 1.49 & 0.11 & 115.2 & 2.02 & 8.75 & 15.5 & 1.97 & 0.64 & 1.39 & 144 \\  
C36 & 343.29 & 11.97 & 1.06 & 0.11 & 104.59 & 2.99 & 16.89 & 15.21 & 0.96 & 0.31 & 2.59 & 201 \\  
C37 & 331.16 & 7.62 & 1.55 & 0.11 & 86.19 & 1.99 & 11.46 & 15.52 & 0.85 & 0.27 & 2.84 & 539 \\  
C38 & 334.0 & 55.86 & 0.58 & 0.08 & 98.33 & 3.82 & 23.13 & 14.86 & 0.21 & 0.07 & 1.44 & 11 \\  
C39 & 303.07 & 1.63 & 2.12 & 0.15 & 83.2 & 1.98 & 11.59 & 15.59 & 1.94 & 0.62 & 1.68 & 309 \\  
C40 & 302.16 & -1.21 & 2.03 & 0.17 & 64.45 & 2.42 & 16.38 & 15.69 & 2.28 & 0.74 & 1.99 & 264 \\  
C41 & 257.34 & -0.48 & 0.8 & 0.12 & 17.56 & 4.13 & 20.24 & 15.01 & 1.87 & 0.6 & 2.5 & 116 \\  
C42 & 245.49 & -0.11 & 0.35 & 0.08 & 76.74 & 6.63 & 30.2 & 14.55 & 0.8 & 0.26 & 2.61 & 88 \\  
C43 & 273.17 & -0.82 & 1.58 & 0.17 & 110.77 & 3.1 & 20.49 & 14.05 & 1.43 & 0.46 & 1.71 & 69 \\  
C44 & 305.47 & 1.93 & 0.99 & 0.18 & 140.57 & 5.19 & ~ & 13.93 & 1.08 & 0.35 & 0.5 & 1 \\  
C45 & 21.81 & 0.71 & 1.53 & 0.15 & 25.27 & 2.84 & 31.16 & 15.81 & 3.15 & 1.02 & 1.33 & 97 \\  
C46 & 44.21 & 2.66 & 1.38 & 0.21 & 79.04 & 4.47 & 20.5 & 13.92 & 2.15 & 0.69 & 1.62 & 36 \\  
C47 & 320.49 & -1.23 & 2.82 & 0.13 & 95.19 & 1.29 & 7.43 & 14.63 & 1.98 & 0.64 & 1.76 & 240 \\  
C50 & 20.27 & 1.04 & 2.05 & 0.15 & 132.38 & 2.07 & 14.6 & 15.55 & 2.47 & 0.8 & 1.55 & 255 \\  
C52 & 318.76 & 2.78 & 1.25 & 0.1 & 97.68 & 2.27 & 10.11 & 13.63 & 1.3 & 0.42 & 1.75 & 177 \\  
C53 & 359.17 & 4.97 & 3.56 & 0.32 & 32.58 & 2.59 & 6.4 & 13.42 & 2.3 & 0.74 & 0.91 & 11 \\  
C54 & 305.17 & 1.31 & 1.53 & 0.11 & 75.46 & 1.97 & 15.25 & 14.37 & 1.33 & 0.43 & 1.34 & 137 \\  
C55 & 15.15 & 1.68 & 1.08 & 0.12 & 126.81 & 3.26 & 19.57 & 15.22 & 2.3 & 0.74 & 1.89 & 278 \\  
C56 & 14.97 & -0.96 & 0.55 & 0.11 & 126.17 & 5.84 & 36.34 & 14.43 & 1.41 & 0.46 & 1.29 & 55 \\  
C57 & 40.59 & 4.14 & 1.75 & 0.13 & 142.0 & 2.1 & 10.57 & 16.4 & 2.81 & 0.91 & 2.18 & 343 \\  
C58 & 351.32 & 0.6 & 1.26 & 0.11 & 54.96 & 2.62 & 22.59 & 15.74 & 2.48 & 0.8 & 1.3 & 193 \\  
C61 & 0.48 & 2.19 & 4.08 & 0.12 & 49.25 & 0.84 & 6.7 & 15.82 & 2.4 & 0.77 & 1.44 & 295 \\  
\enddata
\tablecomments{Main properties of the filtered IPS-GI subsample, mean values per field. The columns are: 1) Field number; 2) Galactic Longitude; 3) Galactic Latitude; 4) Degree of Polarization (weighted); 5) Error on Degree of Polarization (weighted); 6) Polarization angle in Galactic coordinates (weighted); \edit2{7) Error on Polarization angle (weighted); 8) Dispersion (circular standard deviation) of the Galactic polarization angles in the field;} 9) V band magnitude; 10) V band extinction taken from \citet{anders2022photo}; 11) Reddening (Using $E(B-V) = A_{V}/3.1$); 12) Distance taken from \citet{anders2022photo}; 13) Number of stars in field.}
\tablenotetext{a}{The letter C refers to the word \emph{Campo}, which is Portuguese for field.}
\end{deluxetable*}

\clearpage

\section{Correlations between stellar parameters}
\label{sec:corr}
\subsection{Polarimetry and spatial parameters \label{corr:pol_spat}}
Due to the nature of the polarization, we expect to see correlations between certain parameters. For example, a decrease of the polarization for higher Galactic latitudes due to the decrease in integrated dust content along those lines of sight. To investigate this as well as other correlations, the data was divided into nearby ($d < 4$ kpc) and distant ($d > 4$ kpc), as well as high ($|b| > 15\degr$) and low ($|b| < 15\degr$) latitude. The weighted average \edit1{degrees of polarization and average Galactic polarization angles for these bins are} presented in Table \ref{tab:binnedata}. The values were calculated following the method of \citet{pereyra2007polarimetry}. This first coarse look appeared to reveal two expected correlations. Firstly, the polarization increases as a function of distance. We expected to see this, because the polarization of the starlight is caused by dichroic extinction in the ISM. As the starlight travels through more ISM dust, the light becomes more polarized if the magnetic field projection does not change considerably along the line of sight. In the low latitude subsample of stars, the average polarization increases by \edit2{0.7} percentage point when comparing nearby and distant stars. 

\begin{deluxetable*}{l l l l l }
\tablenum{2}
\tablecaption{Polarimetric averages at different latitudes and distances.\label{tab:binnedata}}
\tablewidth{0pt}
\tablehead{
\colhead{Lat.} & \colhead{Dist.} & \colhead{N Stars (\%)} & \colhead{P} & \colhead{$\theta_{Gal}$} \\
\colhead{} & \colhead{} & \colhead{} & \colhead{(\%)} & \colhead{(deg)}
}
\decimalcolnumbers
\startdata
         & Total    & 20 (100)      & $0.44 \pm 0.10$   &  $98.04 \pm 6.22$   \\
High     & Nearby   & 19 (95)      & $0.44 \pm 0.09$   &  $97.46 \pm 6.42$   \\
         & Distant  & 1 (5)       & $4.87 \pm 0.47$   & $129.18 \pm 2.76$  \\ \hline
         & Total    & 10496 (100) & $1.39 \pm 0.13$   & $89.94 \pm 1.73$  \\
Low      & Nearby   & 9272 (88)   & $1.37 \pm 0.12$   & $89.19 \pm 2.53$  \\
         & Distant  & 1224 (12)    & $2.07 \pm 0.21$   & $101.62 \pm 2.91$  
\\ \hline
\enddata
\tablecomments{Weighted average degree of polarization and angle of polarization and their associated errors for several bins. High and low latitude stars are located at $|b| > 15\degr$ and $|b| < 15\degr$, respectively. Distant and nearby stars have $d > 4$ kpc and $d < 4$ kpc, respectively.}
\end{deluxetable*}

Secondly, high latitude stars show less polarization compared to low latitude stars. The starlight from low latitude stars passes through more dust than their high latitude counterparts, thus leading to an increase in average polarization. \edit2{Indeed, we find that the V-band extinction in the low-latitude sample is much higher than in the high-latitude sample: $A_{V}=1.7\textrm{ mag}$ for the low-latitude sources versus $A_{V}=0.9\textrm{ mag}$ for the high-latitude stars.} Comparing low and high latitude nearby stars, the degree of polarization increases by \edit2{almost a full} percentage point. However, it remains important to acknowledge the low number of stars in the high latitude subsample.

\par
\begin{figure*}[ht!]
\plottwo{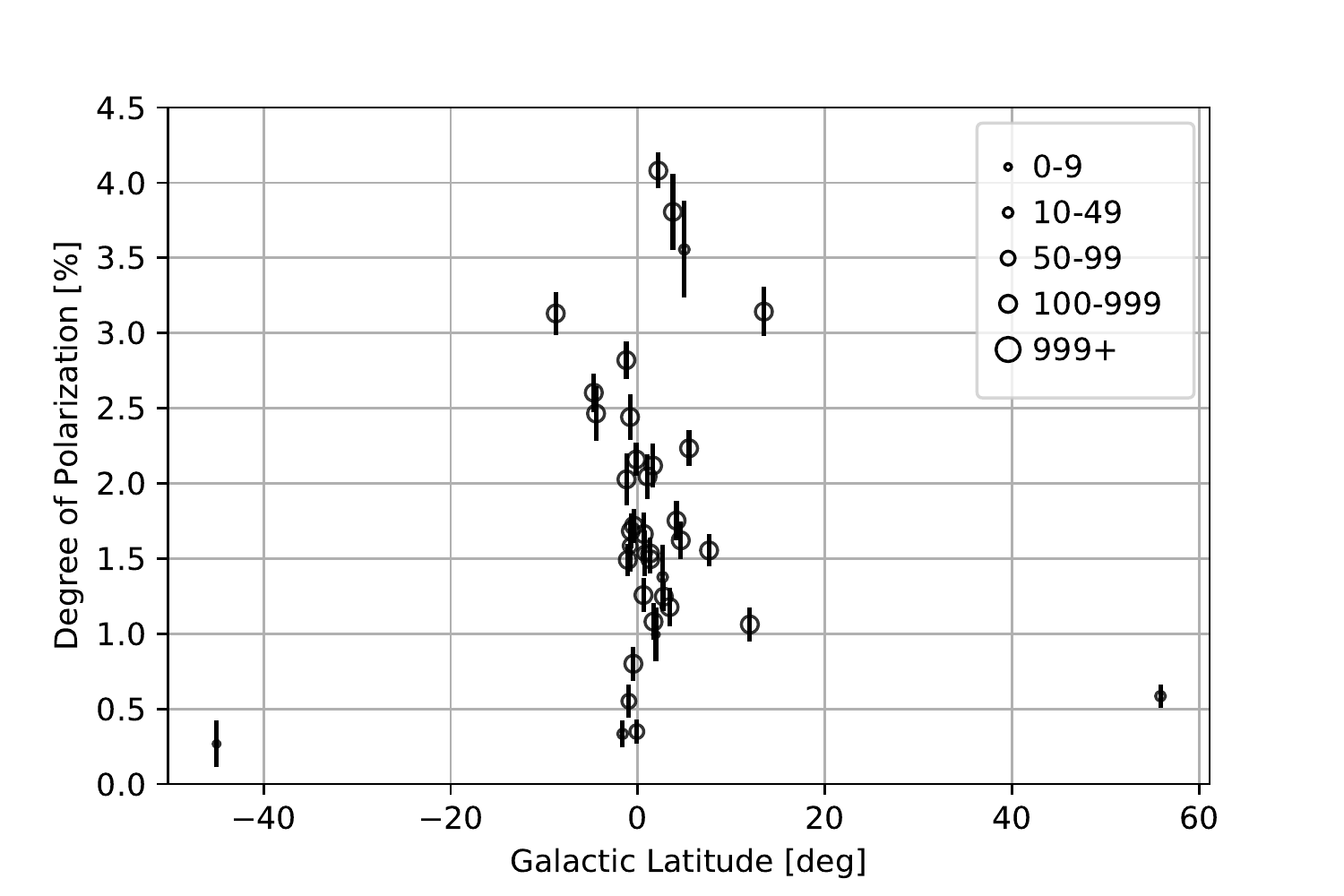}{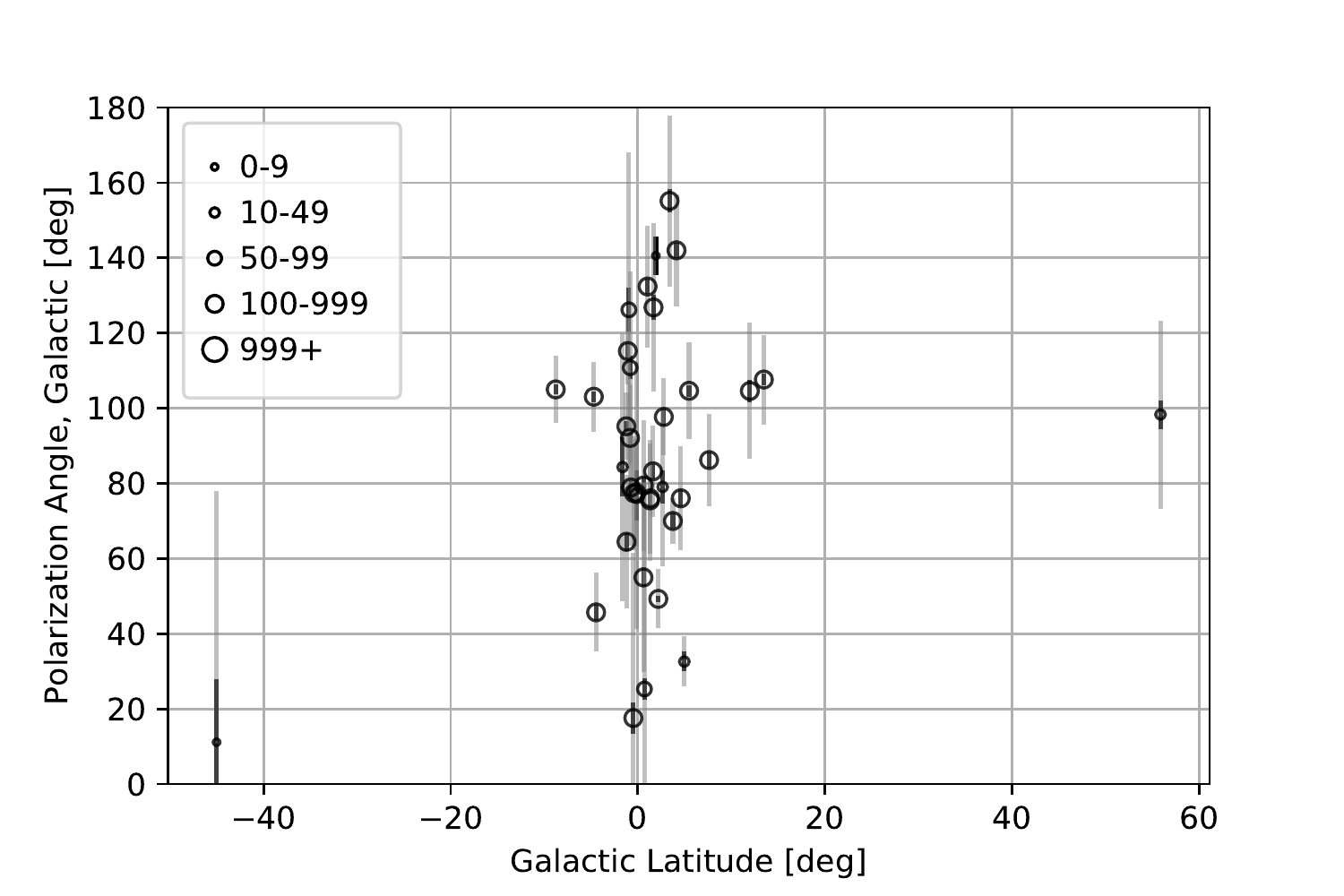}
\caption{Average polarimetric measurements per field \edit2{as a function of Galactic latitude}. Marker sizes scale with the number of stars in that field. Left:  weighted average degrees of polarization \emph{P} in percentage per field. Error bars represent the weighted error. Right: average polarization angles $\theta$ in degrees per field. Gray error bars denote the standard deviation of the distribution within each field. The black error bars represent the average measurement error within each field.}
\label{fig:latbins}
\end{figure*}
\begin{figure*}[ht!]
\plottwo{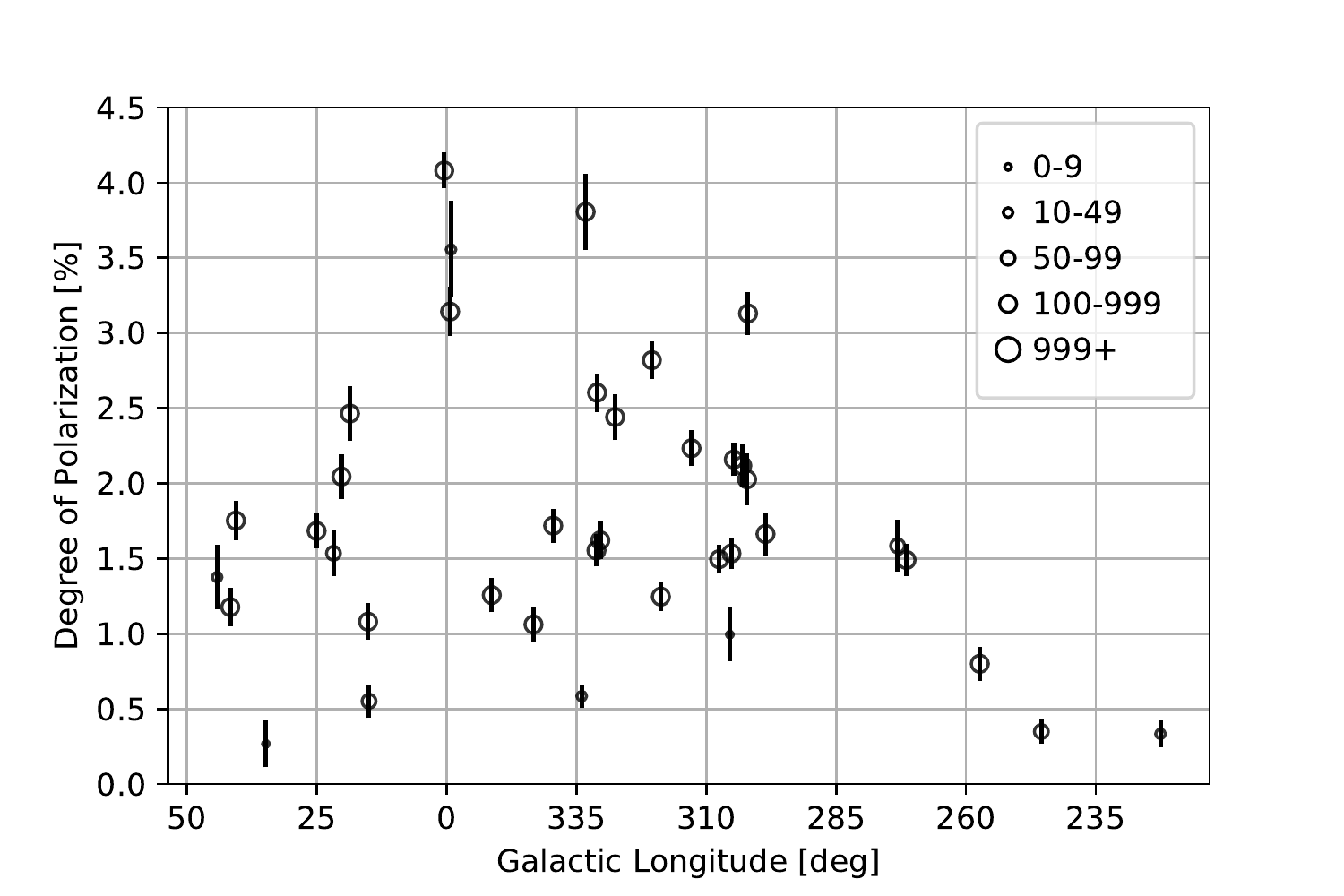}{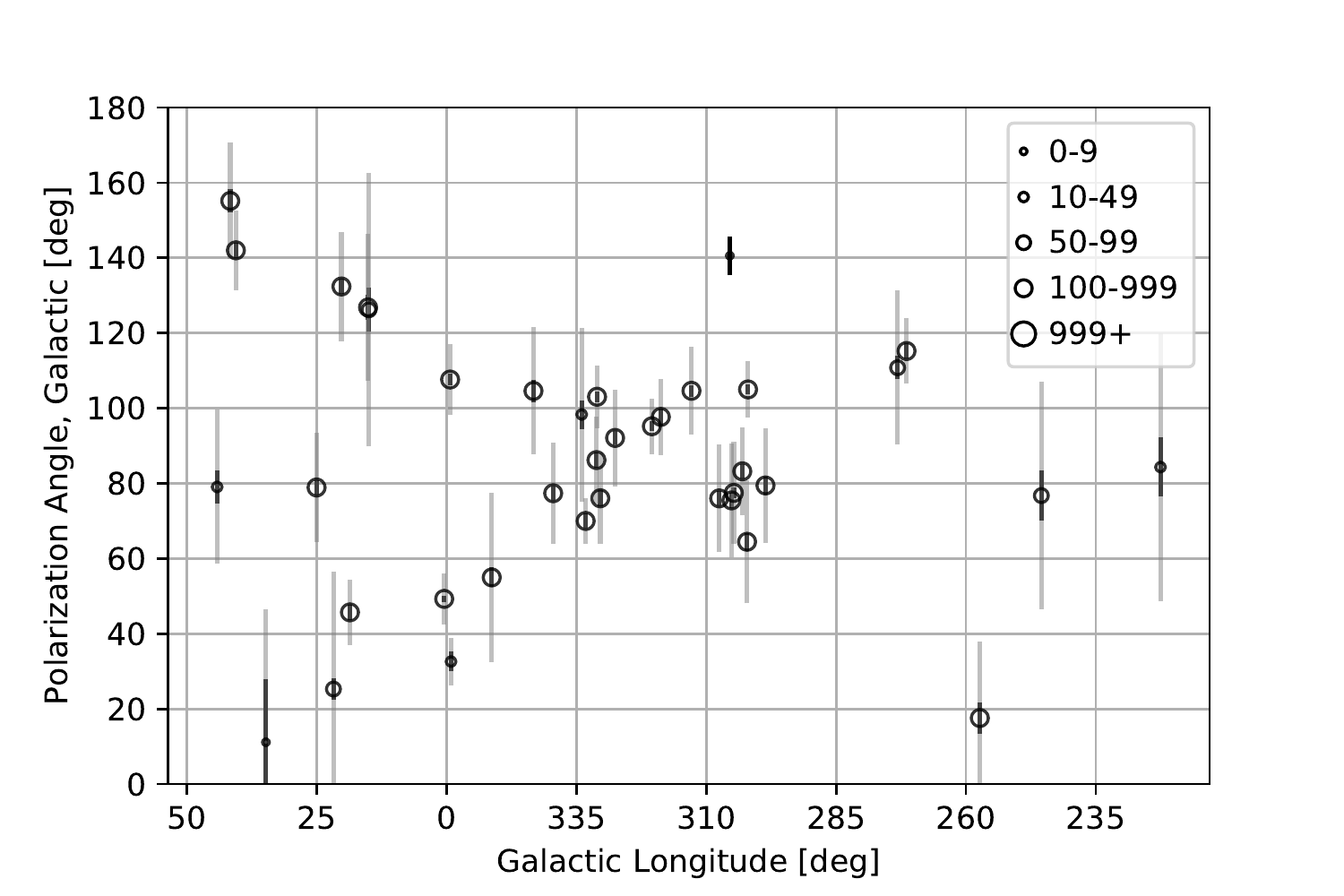}
\caption{Average polarimetric measurements per field \edit2{as a function of Galactic longitude}. Marker sizes scale with the number of stars in that field. Left:  weighted average degrees of polarization \emph{P} in percentage per field. Error bars represent the weighted error. Right: average polarization angles $\theta$ in degrees per field. Gray error bars denote the standard deviation of the distribution within each field. The black error bars represent the average measurement error within each field.}
\label{fig:lonbins}
\end{figure*}

We next present the distributions of Galactic latitude and longitude in conjunction with polarimetric parameters P and $\theta$ on a field-by-field basis, revealing a rich structure, as can be seen in Figures \ref{fig:latbins} and \ref{fig:lonbins}.
Firstly, looking at latitude confirmed the findings from Table \ref{tab:binnedata}. As can be seen in Figure \ref{fig:latbins}, left, the highest weighted average degrees of polarization are found in fields located near the center of the Galactic plane at b = 0\degr. As for the polarization angle in Figure \ref{fig:latbins}, right, we note that the fields around the b = 0\degr show a wide range of polarization angles centered around $\theta_{Gal} \approx 90\degr$, whereas the two bins \edit2{above and below the Galactic plane show very low polarization. We note, however, that the fields further away from the plane are sparsely populated. }

\par
We also investigated average values by longitude. Figure \ref{fig:lonbins}, left shows the weighted average degrees of polarization per field. The polarization appears to be highest in the direction of the Galactic center and its immediate surroundings. This is caused by the higher densities of dust associated with that region \citep[see for example][]{rezaei2017inferring, misiriotis2006distribution}. Plotting the average polarization angle per field as in Figure \ref{fig:lonbins}, right, reveals an average polarization angle $\theta_{Gal} \sim 90\degr$ throughout the longitude range, with significant outliers above and below that value, see also the vectors in Fig. \ref{fig:skyplot}. As described in Section \ref{avgperfield} and Table \ref{tab:perfield}, these fields may be indicative of variations of the magnetic field across the Galactic plane or intervening structures such as spiral arms throughout the line of sight.

\begin{figure*}[ht!]
\plottwo{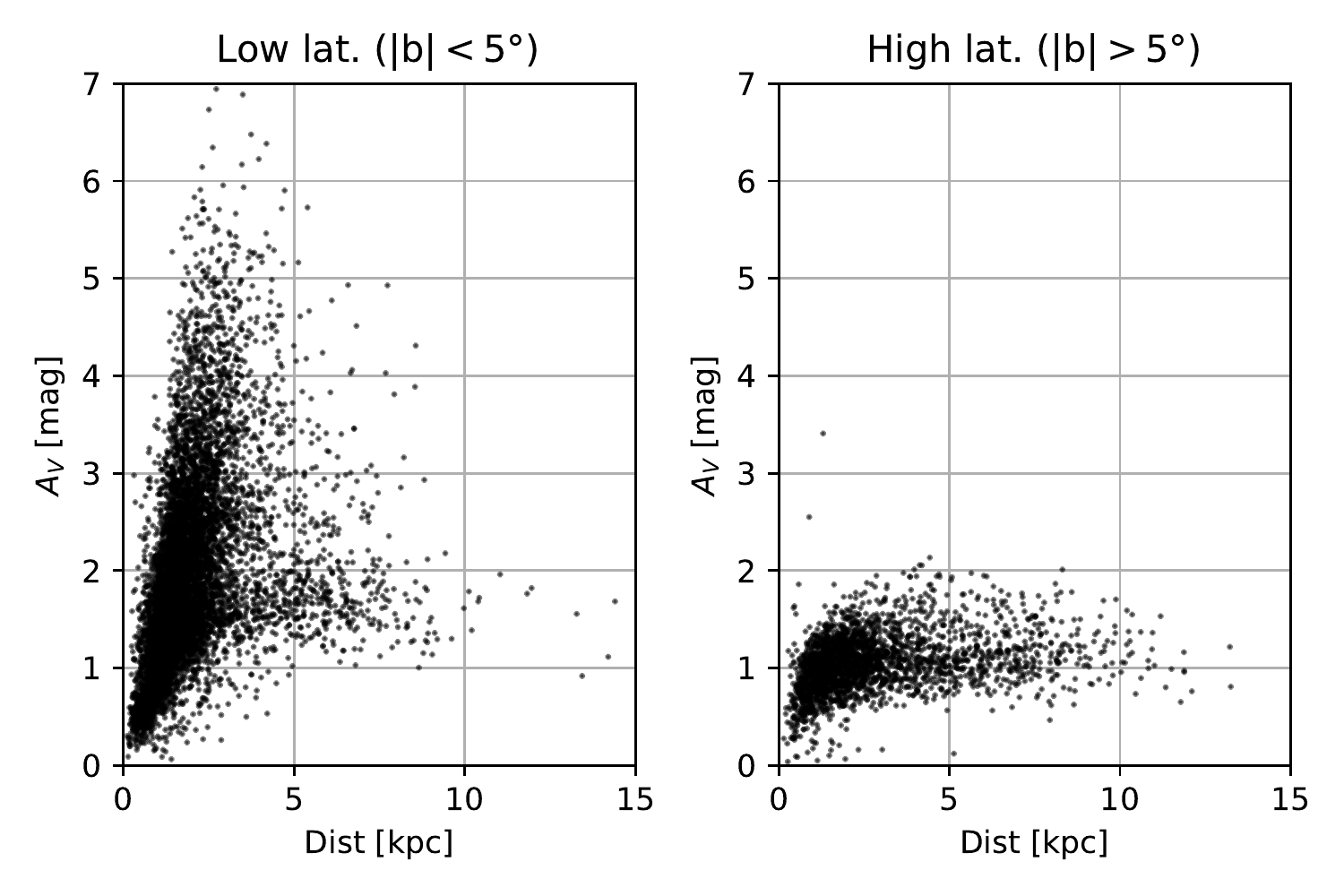}{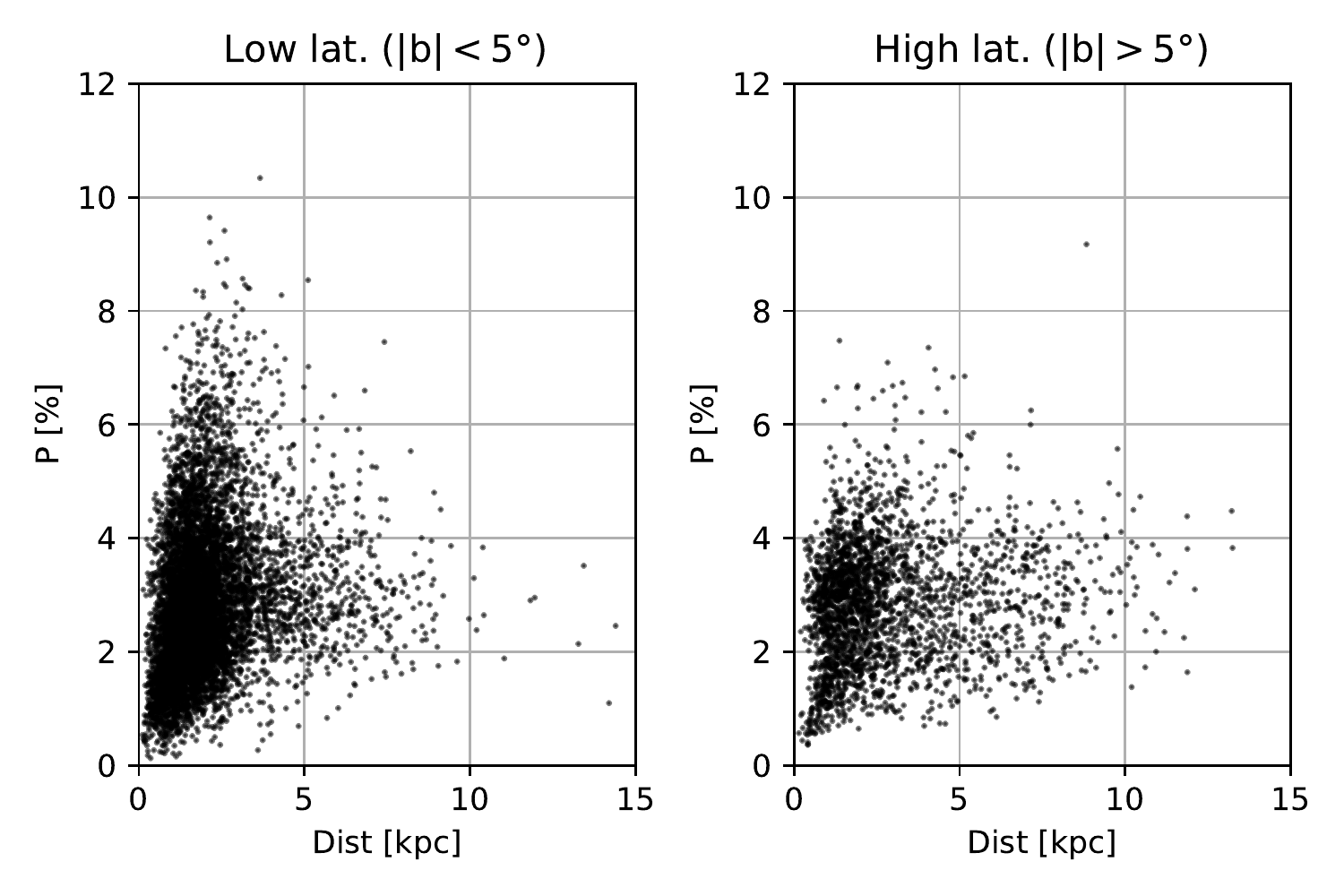}
\caption{Left: V-band extinction $A_V$ versus distance for low-latitude ($|b|<5\degr$) and high-latitude ($|b|>5\degr$) stars. Right: degree of polarization versus distance for low-latitude ($|b|<5\degr$) and high-latitude ($|b|>5\degr$) stars. Distances taken from \citet[][]{anders2022photo}}
\label{fig:av_p_dist}
\end{figure*}

\edit2{Figure \ref{fig:av_p_dist} shows the V-band extinction $\textrm{A}_V$ and degree of polarization P as a function of distance, for the low latitude ($|\textrm{b}|<5\degr$) and high latitude ($|\textrm{b}|>5\degr$) stars. As can be seen in the Fig. \ref{fig:av_p_dist}, left, at low latitudes, i.e. in the Galactic plane, we are able to trace dust out to distances of at least 3 kpc. The continuous increase in extinction indicates a significant dust presence out to large distances. The case for the high latitude stars is very different, where the V-band extinction flattens out beyond a distance of 1 kpc, beyond which we are less sensitive to dust and, by extension, polarization. The relationship between polarization and distance is more complicated, see also Fig. \ref{fig:av_p_dist}, right. Although we are able to trace dust out to large distances, the degree of polarization does not necessarily increase linearly with distance. }

\par
\begin{figure*}[ht!]
\plottwo{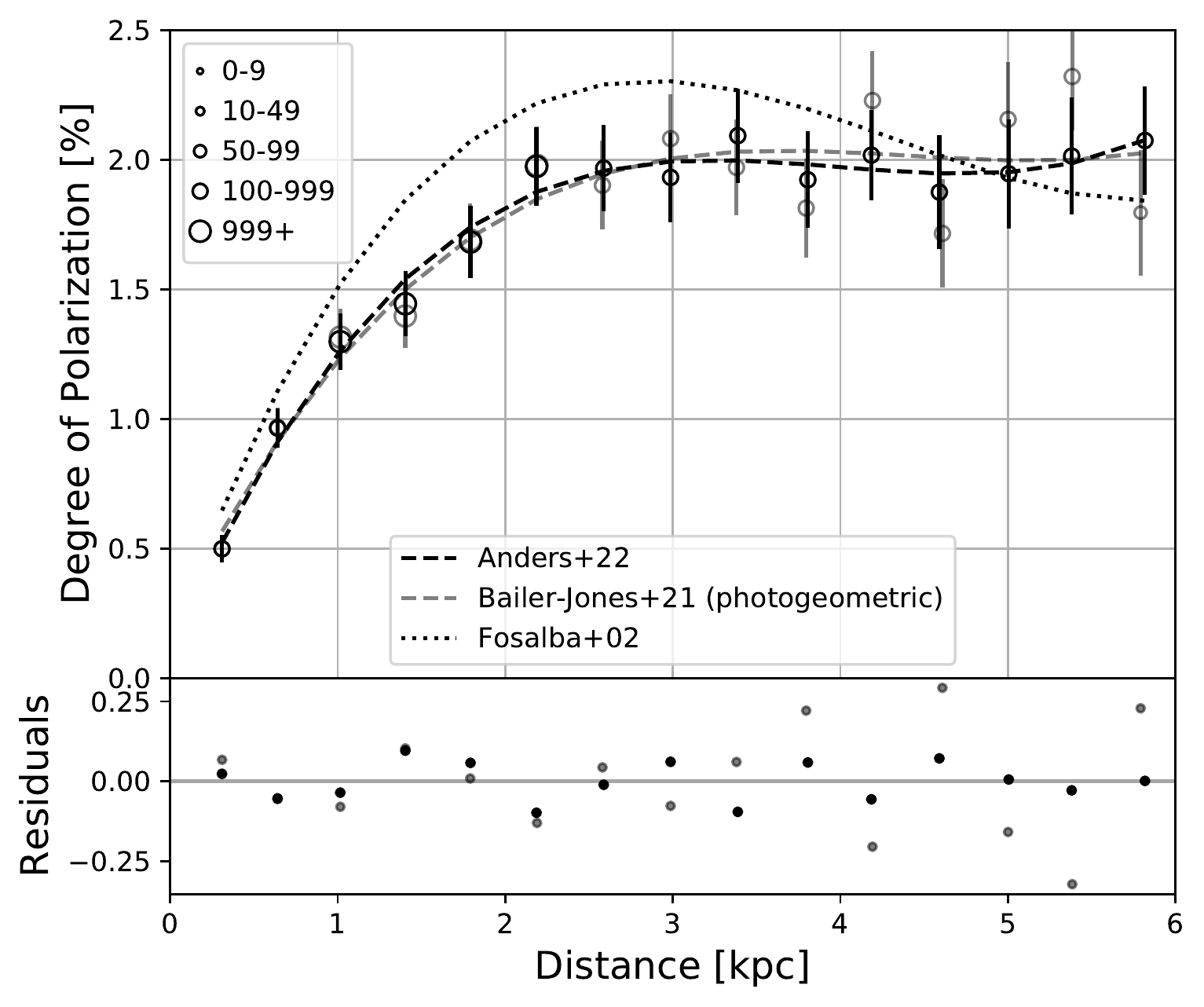}{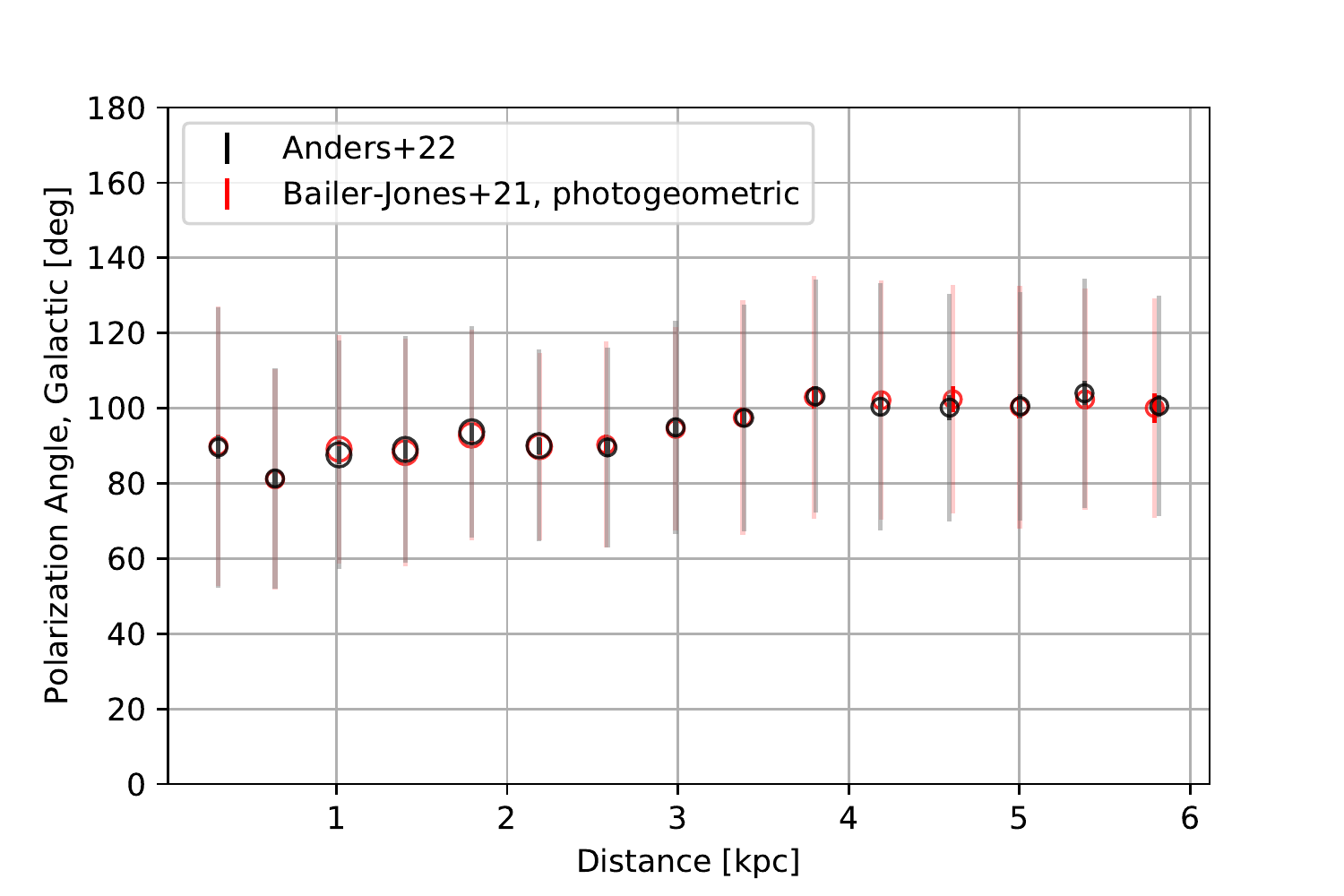}
\caption{Average polarimetric measurements per 400 pc distance bin. \edit2{Plots have been limited to $d\le6 \textrm{ kpc}$ to ensure each bin has a significant population of stars.} Marker sizes scale with the number of stars in that bin. \edit2{Left: weighted average degree of polarization \emph{P} in percentage per distance bin. Error bars represent the weighted error. Dashed black line data \citep{anders2022photo} and the gray dashed line \citep[][]{bailer2021estimating}, photogeometric, denote a third degree polynomial fit to the data. The dotted line denotes the \citet{fosalba2002statistical} fit to data from Heiles' agglomeration. The residuals show the difference between the data points and each dashed line, colored as described above.} Right: weighted average polarization angles $\theta$ in degrees per distance bin. \edit2{Black and red markers represent distances taken from \citet[]{anders2022photo,bailer2021estimating}, respectively. The dark (smaller) error bars represent the average measurement error within each bin. Light gray and light red error bars indicate the standard deviation in polarization angles in that bin.}}
\label{fig:distbins}
\end{figure*}

\edit2{We analyzed polarimetric parameters and distance in more detail}. Firstly, we binned the data in bins of 400 pc\edit2{, limited to a maximum distance of 6 kpc to ensure each bin has a significant population of stars}. Figure \ref{fig:distbins} shows the relevant distributions. Figure \ref{fig:distbins}, right, shows a stable average position angle $\theta_{Gal} \sim 90\degr$ out to $d \sim 6 \mathrm{kpc}$. This is indicative of a dominant ordered magnetic field. \edit2{The position angle $\theta_{Gal}$ gives information on the magnetic field along the line of sight out to the distance where there is still a significant amount of dust present. The flattening beyond 3 kpc may be explained by, for example, a change in the large-scale magnetic field leading to depolarization as the signal is integrated along the line-of-sight}. Another possibility is a lack of dust beyond a certain distance. As light passes through magnetic field-aligned dust, we expect the degree of polarization to increase. As we go beyond the dusty thick Galactic disc, we expect the dust content to decrease dramatically, indicated by a flattening of the increase of polarization. However, the precise profile will depend on the line of sight. \edit2{Bias effects due to incompleteness of (weaker) sources at larger distances and/or signal-to-noise ratio filtering may slightly increase the degree of polarization. However, analysis of this is beyond the scope of the paper.}

To further investigate the correlation between distance and degree of polarization, we applied a polynomial fit to the data presented in Figure \ref{fig:distbins}, left. The overall trend appears to agree well with a third-degree polynomial. A fit to the data \edit2{using the distances taken from \citet[]{anders2022photo}} finds the following coefficients

\begin{equation}
\label{eq:3rdegfit}
    P = 0.087 + 1.526d - 0.396d^{2} + 0.033d^{3}
\end{equation}

\noindent where P is the degree of polarization in percent and d is the distance in kiloparsec. The bottom of Figure \ref{fig:distbins}, left show the residuals for each data point. The third degree polynomial trend is in agreement with findings from \citet{fosalba2002statistical} based on \citet{heiles20009286} data, see also the dotted line in Fig. \ref{fig:distbins}, left, although the coefficients found therein differ slightly from those in Eq. \ref{eq:3rdegfit}. (\citet{fosalba2002statistical} find \begin{math} P = 0.13 + 1.81d - 0.47d^2 + 0.036d^3 \end{math}, see Eq. 1 therein). We note that there is no physical basis for applying a third-degree polynomial and physical information cannot be derived from the coefficients of the fit. Furthermore, we emphasize that the distribution across the sky of stars and the depth of the survey differ greatly between \citet{fosalba2002statistical} and the IPS-GI. Nonetheless, it is remarkable that we independently found an overall trend that is very similar to \citet{fosalba2002statistical}. \edit2{We find similar trends regardless of distance catalog used. The polynomial curve fitted to the \citet[]{bailer2021estimating} distances is in agreement with Eq. \ref{eq:3rdegfit} within the errors on the parameters.}

\subsection{Polarimetry and photometric parameters \label{corr:pol_phot}}

\begin{figure*}[ht!]
\plottwo{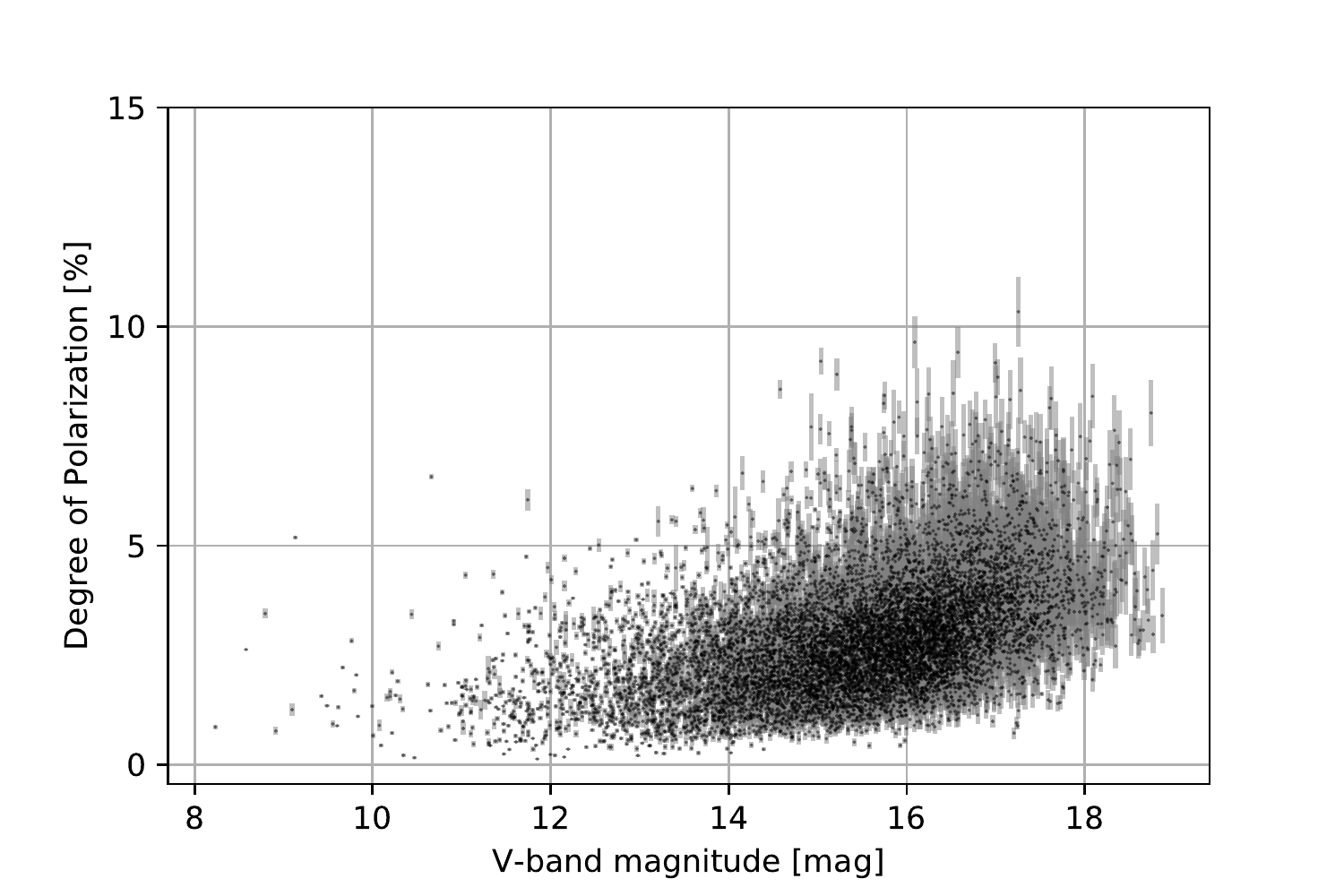}{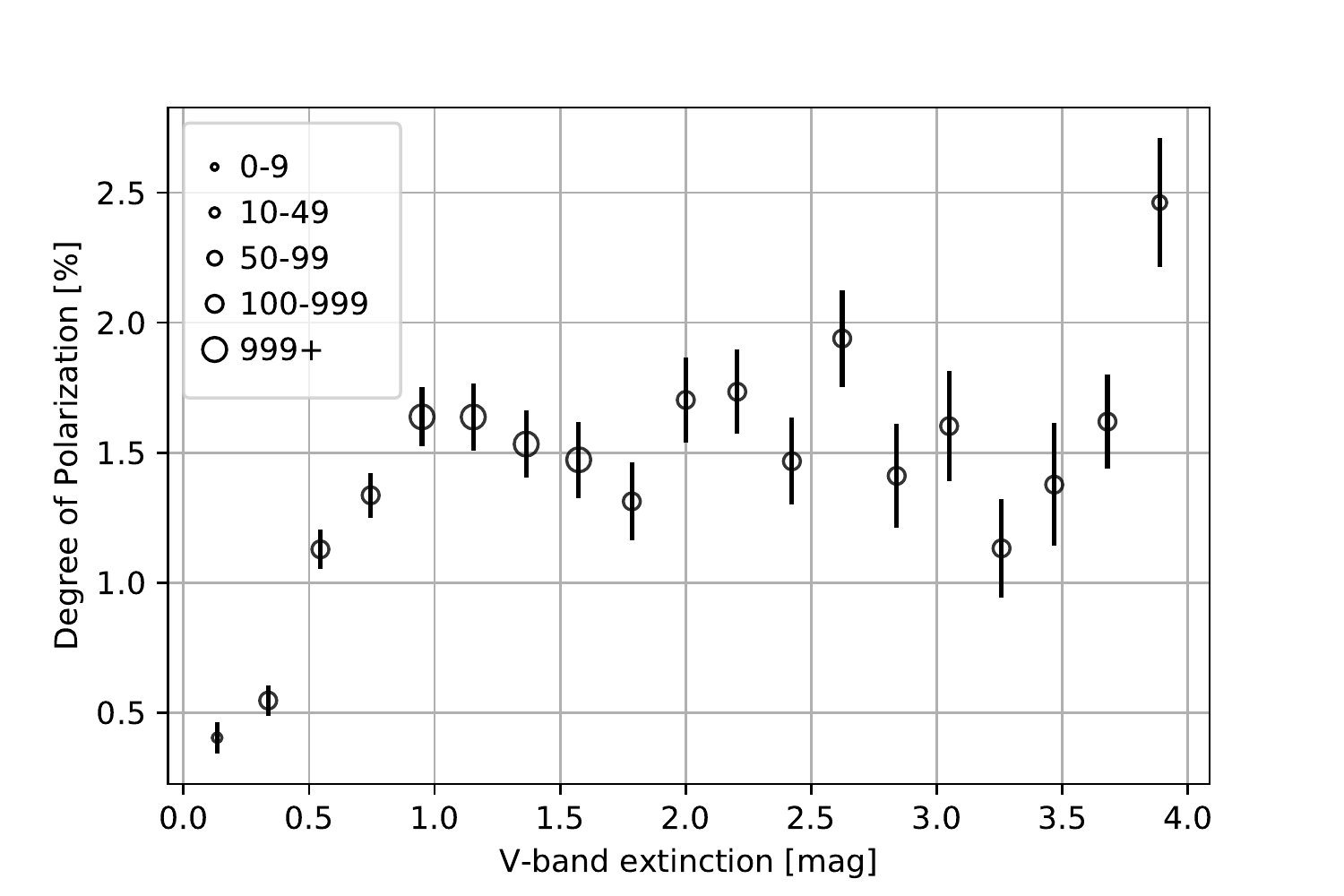}
\caption{Left: degrees of polarization \emph{P} in percentage versus V-band magnitude. The error bars represent the measurement error. Right: weighted average degrees of polarization \emph{P} in degrees per V-band extinction bin. \edit2{Plots have been limited to $A_V<4\textrm{ mag}$ to ensure each bin has a significant population of stars.} Extinctions taken from \citet{anders2022photo}. The error bars represent the average measurement error within each bin. Marker sizes scale with the number of stars in that bin.}
\label{fig:pol_phot}
\end{figure*}

Next, we investigated the correlations between degree of polarization and various photometric parameters. Figure \ref{fig:pol_phot}, left, shows the degree of polarization as a function of the V-band magnitude for the entire sample of stars. Although the density of stars increases greatly towards higher magnitude, peaking at V $\sim$ 16, two trends are apparent. Firstly, we note the increase of the error in the degree of polarization towards higher magnitude. Fainter stars appear to have a larger error in the degree of polarization, which is expected. Despite employing longer observing times to include higher magnitude stars in the sample, a higher error cannot fully be eliminated. Secondly, the degree of polarization itself appears to increase with higher magnitude. A similar trend is visible in the \citet{heiles20009286} data, although it must be noted that the highest magnitude therein is V $\sim$ 13. It is important to note that in this case, the lack of lower polarization, higher magnitude sources is partially caused by the filtering, especially the ${P}/\sigma_{P}$ filter that excludes low polarization stars. As the error in polarization is relatively higher for those sources, they are inadvertently filtered out. Finally, stars observed under less than ideal weather conditions will tend to have larger errors due to, for example, a lower photon count. They would contribute to the spread in accuracy beyond what could be expected from the (unattenuated) magnitudes.
\par
Figure \ref{fig:pol_phot}, right shows the average polarization per V-band extinction (A\textsubscript{V}, taken from \edit1{ \citealt{anders2022photo}}) bin. The data was binned to extinction bins of 0.21 mag wide. As the extinction is a reliable tracer of the dust contents along a line of sight, the apparent increase of polarization as a function of that extinction is expected. \edit2{Fig. \ref{fig:av_p_dist}, left, shows that for the lower-latitude fields specifically, we are able to trace dust to large distances of at least 3 kpc, as is indicated by the continuous increase in extinction with distance shown in the left-most plot. Although this does not necessarily correspond to an increase in polarization (e.g. depolarization may occur along a line-of-sight), we are able to probe the larger distances in terms of polarimetry.} A detailed analysis of the polarization behavior with extinction will be presented in a forthcoming paper (Angarita et al., in prep).

\section{Discussion}
\label{sec:disc}
We presented a first look at the IPS-GI data, which contains data of \edit2{over 40000} linearly polarized stars in 38 fields distributed over the Southern Sky, mainly close to the Galactic plane that sample the diffuse ISM. \edit2{We've added distances and photometric data from auxiliary catalogs and} analyzed a high-quality subsample of 10516 stars. The degree and position angle of the linear interstellar polarization depend on the quantity and alignment properties of the dust along the line of sight. The polarization angle denotes the line-of-sight averaged orientation of the magnetic field component in the plane of the sky. As the polarization is not a scalar, the effect of more than one dust cloud or other intervening structure is not necessarily additive in the polarization degree.
\par
\edit2{\subsection{Degree of Polarization as a function of longitude and latitude}}

\edit2{The degree of polarization depends on the dust distribution, dust alignment and magnetic field direction. Dust rotational alignment with the magnetic field is thought to be close to 100\% in diffuse gas \citep[see e.g.][]{mathis1986alignment, kim1995size, panopoulou2019extreme}. Considering only the magnetic field orientation, we expect the degree of polarization to be maximal when the magnetic field is oriented in the plane of the sky, and minimal when the magnetic field is along the line of sight. For a uniform magnetic field following the spiral arms, this would produce a maximum degree of polarization in the direction of the Galactic Center and decreasing polarization degrees away from the Galactic Center. Our data is consistent with this expectation, see Fig. \ref{fig:lonbins}, left. However, Fig. \ref{fig:lonbins}, left, also displays large variations in polarization degree as a function of longitude, which can be caused either by structures in the magnetic field direction and/or variations in the dust distribution along the line of sight. Interstellar turbulence will partially depolarize the radiation on path lengths longer than a few 100 pc, leading to a decrease in the degree of polarization. In addition, the presence of spiral arms can significantly distort magnetic field directions, possibly inducing a complex longitude dependence in the polarization degree data \citep[see e.g.][]{gomez20043dMHD}.}

The observed distribution of degree of polarization centers around 2\% (Fig \ref{fig:photpol_distribs}, middle) and the average polarization degree is higher than in the \citet{heiles20009286} catalog \citep[][]{fosalba2002statistical}, which is likely because the IPS-GI fields are more concentrated towards \edit2{lower Galactic latitudes} than the Heiles compilation. In addition, the IPS-GI contains more distant stars, which may show a more strongly polarized signal. 

\edit2{Varying dust distributions will induce differences in polarization degree as a function of Galactic latitude and longitude. Most notably, at high latitudes dust will only be present in the nearest part of the line of sight: for a field at $b \sim 45^{\circ}$, a dust layer of 150 pc half-thickness \citep[][]{sparke2007galaxies} would result in dust only along the nearest $\sim200$~pc. This is apparent in the low degrees of polarization in the two high-latitude fields (Fig \ref{fig:latbins}, left). The variable dust distribution in the Galactic disc, combined with varying magnetic field directions, causes the large scatter in polarization degree in Figs. \ref{fig:lonbins}, left and \ref{fig:latbins}, left.}

\edit2{\subsection{Polarization angle as a function of longitude and latitude}}

The observed distribution of polarization angle is surprisingly uniform over each 0.3x0.3deg field, with a clear preferred angle for most of the fields (as for example in Fig \ref{fig:campo_47_example}). This is a clear indication that the turbulent component of the magnetic field, thought to have maximum correlation scales up to a few $\sim100$ pc \citep[][]{beck2016new} is averaged out along the (\edit2{$\sim$}kpc) lines of sight towards the IPS-GI stars \edit2{at low latitudes}.

\edit2{For the two fields at the high latitudes, where significant dust only exists in the first few 100 pc along a line of sight, contributions from the turbulent magnetic field component will play a more significant role. For fields at intermediate latitudes ($|b|\sim~10^{\circ}$), a uniform dust layer of $\sim150$~pc half-thickness would end at $\sim 850$~pc distance, which means that some polarization angle scatter due to turbulence may still be there. Closer to the plane, with longer path lengths containing dust, variations in the mean polarization angle are the largest (see Fig \ref{fig:lonbins}, right), although interstellar turbulence is expected to play a smaller role in that region due to the long path lengths involved. Variations in mean polarization angle are created by a changing direction of the plane-of-sky component of the magnetic field, weighted by the dust distribution. These variations can be either due to larger-scale structures in the magnetic field such as spiral arms \citep[][]{gomez20043dMHD} or small-scale structures such as dust clouds along the line-of-sight. In addition, the number of inhomogeneities in the dust distribution is higher close to the Galactic plane, causing additional variations in mean polarization angle per field. Even though the IPS-GI fields were selected to not have major dense structures along their lines-of-sight, there are still a number of fields for which dust clouds dominate the polarization signal along the line of sight (Versteeg et al, in prep). }

\par
Furthermore, in most fields the field-averaged polarization angle has the same orientation as the magnetic field orientation averaged over the whole line of sight through the Milky Way as inferred by Planck polarized dust data (\edit2{Figs. \ref{fig:skyplot}, \ref{fig:plank_comp}}). As the IPS-GI stars only probe part of this line of sight, the similarity of angle orientation between IPS-GI and Planck points to a remarkable uniformity of the large-scale magnetic field direction along these lines of sight. However, there are exceptions to this uniformity in polarization angles, as will be discussed in a forthcoming paper.
\

\begin{figure*}[ht!]
\plottwo{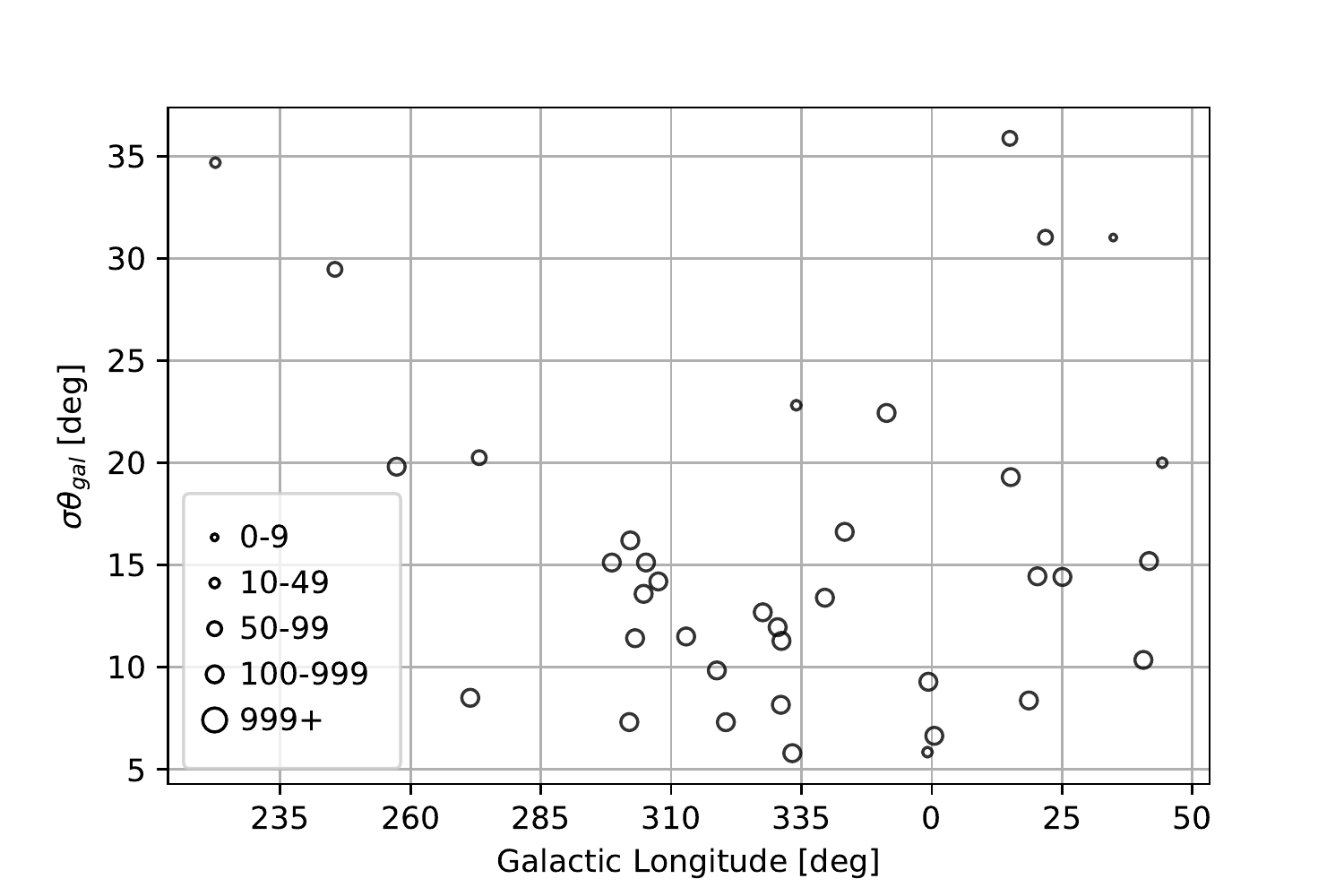}{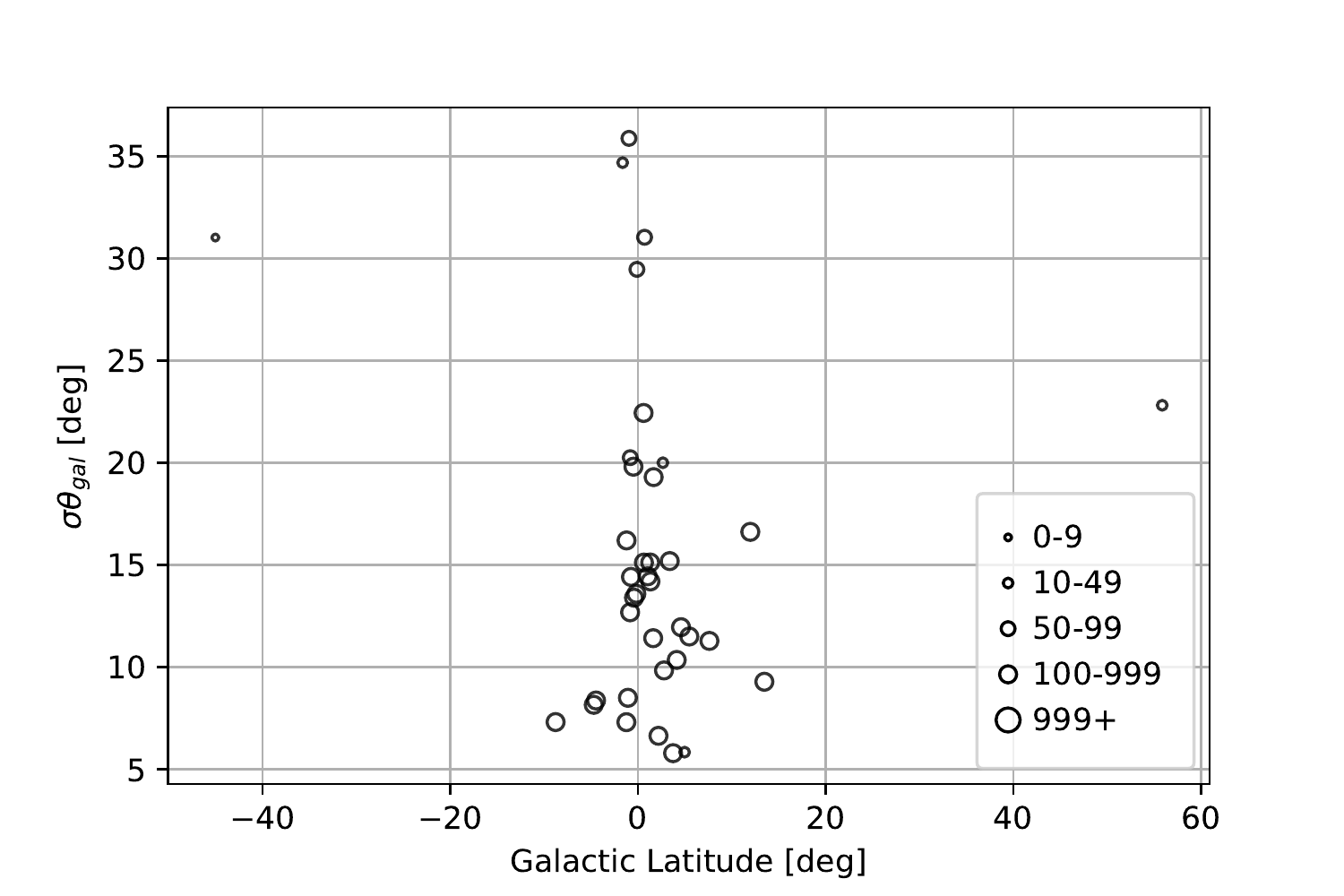}
\caption{Left: polarization angle dispersion $\sigma\theta_{Gal}$ in degrees versus Galactic longitude. Right: polarization angle dispersion $\sigma\theta_{Gal}$ in degrees versus Galactic latitude. Marker sizes scale with the number of stars in that bin.}
\label{fig:dispersions}
\end{figure*}

\edit2{\subsection{Polarization angle dispersions}}
\edit2{The dispersion in polarization angle is presented in Table \ref{tab:perfield} and its dependence on Galactic longitude and latitude as shown in Figure \ref{fig:dispersions}. Most angle dispersions are between 5 and 25 degrees, except for a few fields. Firstly, the two high-latitude fields contain so few stars that the calculated angle dispersion is very unreliable. Secondly, there are 5 fields with high angle dispersions around 30 degrees (C30, C34, C42, C45 and C56), four of which are located close to the Galactic plane. Field C34 is located at high latitude, but contains only 9 stars. All fields are characterized by a relatively low number of stars, and fields C30, C42 and C56 also show deviating distributions of polarization degree (Angarita et al., in prep). Therefore, it is possible that these fields contain anomalous structures (for example, C56 is located at the edge of an HII region), or have deviating dust and/or magnetic field properties. However, it should be noted that \citet[]{medan2019lbw} note a similar dependence of angle dispersion as a function of Galactic longitude in their study of polarized starlight from magnetized dust in the Local Bubble wall, which they suggest might be connected to the presence of OB associations changing the radiative alignment of the grains.}

\edit2{\subsection{General remarks}}
\edit2{For a uniform distribution of homogeneous dust along the line of sight in a homogeneous magnetic field, one would expect a linear increase degree of polarization with extinction. For nearby stars (up to an extinction of about $\sim 1\textrm{ mag}$), the IPS-GI stars show this behavior on average (Fig \ref{fig:pol_phot}, right). However, at higher extinctions the relation between degree of polarization and extinction flattens out, indicating a lower increase in polarization with extinction due to depolarization. The depolarization is mostly caused by variable orientations along the line of sight of the plane-of-the-sky magnetic field component, due to interstellar turbulence and/or meso-scale structure like spiral arms.} \edit1{We also emphasize that most of the data presented above has distances smaller than 3 kpc. Although high quality data exists at higher distances, the associated average uncertainties also increase. We are therefore less sensitive to structures that may exist beyond 3 kpc.}
\par
\edit1{We note that our} analysis is based on the assumption that the starlight we observed is intrinsically unpolarized. While this may be true for most stars, we cannot eliminate all intrinsically polarized sources through the applied filters. When studying individual stars in this catalog, it is therefore important to keep this in mind. However, considering the uniformity of parameters and their agreement with the assumption that our stars are intrinsically unpolarized, we do not expect intrinsically polarized stars to significantly influence the overall statistics of the catalog. We do note that stars with deviating polarization angle (see for example in Figure \ref{fig:c47_pol_distribs}, right) are candidates for further a more detailed analysis of their polarimetric properties.
\par
In addition, it is of the utmost importance to keep in mind the highly fragmented spatial distribution of sources in the plane of the sky. While it is tempting to use the data presented in this paper to draw generalizing conclusions about the ISM and GMF, this must be done with caution. Due to their small size, 0.3x0.3$\degr$, the fields may not be representative of the general structure and conditions of the surrounding ISM. Any outliers or stand-out features presented in the above sections may be falsely interpreted as characteristics of the whole ISM or the GMF, while it may be one (dominant) field that allows the average values to deviate from a general trend. Therefore, it remains important to compare the data presented here to other starlight polarization sources such as \citet{heiles20009286} as well as complementary data at other wavelengths (e.g. Planck dust polarization \citep{abergel2014planck} as in \edit2{Figs. \ref{fig:skyplot} and \ref{fig:plank_comp}} or GPIPS infrared data \citep{clemens2020galactic}), but this is beyond the scope of this paper. \edit2{The kind of study presented in this paper could benefit from homogeneous data covering a large area on the sky. Planned polarimetric surveys, such as SouthPol \citep[][]{magalhaes2012southpol} and Pasiphae \citep[][]{tassis2018pasiphae}, will contribute greatly to the understanding of the interstellar medium.}

\section{Conclusions}
\label{sec:conc}
We presented a first look at the Interstellar Polarization Survey - General ISM (IPS-GI) catalog. This new catalog contains polarimetric measurements \edit2{for over 40000 stars in 38 distinct fields. We have cross-matched the IPS-GI data to the GAIA EDR3 \citet[][]{gaia2021edr3} catalog, as well as two auxiliary distance catalogs \citet[][]{anders2022photo, bailer2021estimating}}. We applied quality filters, presenting an analysis limited to 10516 of the highest quality observations. We presented various distributions of polarimetric and photometric parameters, as well as correlations between those parameters. As expected, the measured degree of polarization correlates with sky position, as well as distance and V-band extinction as expected. We found evidence of a dominant ordered magnetic field, oriented parallel to the Galactic plane, indicated by a stable Galactic polarization angle $\theta \sim 90\degr$ across different Galactic latitudes, longitudes and distances \edit1{up to 3 kpc}. The measured degree of polarization varies across the sky, although we can associate higher dust contents with a more strongly polarized signal, as expected based on the nature of the polarization. The catalog should enable further research into the structure of the ISM, including a new, more detailed view of the GMF.

\begin{acknowledgements}
\edit2{The authors thank the anonymous reviewer for their insightful comments and suggestions for additions to earlier versions of the manuscript.}

MJFV acknowledges Wouter Veltkamp for his contributions to Figures \ref{fig:campo_47_example} and \ref{fig:skyplot}.

\edit2{Over the years, IPS data has been gathered by a number of dedicated observers, to whom the authors are very grateful: Flaviane C. F. Benedito, Alex Carciofi, Cassia Fernandez, Tibério Ferrari, Livia S. C. A. Ferreira, Viviana S. Gabriel, Aiara Lobo-Gomes, Luciana de Matos, Rocio Melgarejo, Antonio Pereyra, Nadili Ribeiro, Marcelo Rubinho, Daiane B. Seriacopi, Fernando Silva, Rodolfo Valentim and Aline Vidotto.}

MJFV, MH and YAA acknowledge funding from the European Research Council (ERC) under the European Union's Horizon 2020 research and innovation programme (grant agreement No 772663).

AMM's work and Optical/NIR Polarimetry at IAG has been supported over the years by several grants from São Paulo state funding agency FAPESP, especially 01/12589-1 and 10/19694-4. AMM has also been partially supported by Brazilian agency CNPq (grant 310506/2015-8). AMM graduate students have been provided grants over the years from Brazilian agency CAPES.

CVR acknowledges Conselho Nacional de Desenvolvimento Científico e Tecnológico (CNPq), Proc. 303444/2018-5.

This work has made use of data from the European Space Agency (ESA) mission
{\it Gaia} (\url{https://www.cosmos.esa.int/gaia}), processed by the {\it Gaia}
Data Processing and Analysis Consortium (DPAC,
\url{https://www.cosmos.esa.int/web/gaia/dpac/consortium}). Funding for the DPAC
has been provided by national institutions, in particular the institutions
participating in the {\it Gaia} Multilateral Agreement.
\end{acknowledgements}

\edit1{\facility{LNA:BC0.6m}}

\bibliographystyle{aasjournal}
\bibliography{refs.bib}
\end{document}